\documentclass[12pt]{article}
\usepackage[utf8]{inputenc}
\usepackage[scaled]{helvet}
 
\usepackage[T1]{fontenc}

\usepackage{float}
\usepackage[margin = 1 in]{geometry}
\usepackage{amssymb}
\usepackage{amsmath}
\usepackage{graphicx}
\usepackage[table, xcdraw]{xcolor}
\usepackage{hyperref}
\usepackage{subfig}
\usepackage{rotating}
\usepackage{ragged2e}
\usepackage{authblk}
\usepackage{setspace}
\usepackage{multirow}
\usepackage{capt-of} 
\usepackage{url}

\usepackage{xcolor}
\definecolor{bg}{rgb}{0.21, 0.44, 0.57} 
\definecolor{pr}{rgb}{0.47, 0.12, 0.49} 

\usepackage{booktabs}

\usepackage{fancyhdr}
\makeatletter
\renewenvironment{abstract}{%
{\bfseries {\color{bg}\Large\noindent\abstractname\vspace{\z@}}}
\newline}
\makeatother
\usepackage[]{natbib}
\begin{document}

\title{\bf {\color{bg} Characterization of differences in immune responses during bolus and continuous infusion endotoxin challenges using mathematical modeling}}

\author[1]{Kristen A. Windoloski}
\author[2,4]{Susanne Janum}
\author[3,4,5,6]{Ronan M.G. Berg}
\author[1]{\\Mette S. Olufsen}

\affil[1]{\footnotesize Department of Mathematics, North Carolina State University, Raleigh, North Carolina, USA}
\affil[2]{Frederiksberg and Bispebjerg Hospitals, Frederiksberg, Denmark}
\affil[3]{Department of Clinical Physiology and Nuclear Medicine, Copenhagen University Hospital, Denmark}
\affil[4]{Centre for Physical Activity Research, Rigshospitalet, Denmark}
\affil[5]{Department of Biomedical Sciences,  University of Copenhagen, Denmark}
\affil[6]{Neurovascular Research Laboratory, University of South Wales, Pontypridd, UK}

\date{}

\maketitle
\vspace{-0.75cm}
\abstract{\noindent Endotoxin administration is commonly used to study the inflammatory response, and though traditionally given as a bolus injection, it can be administered as a continuous infusion over multiple hours. Several studies hypothesize that the latter better represents the prolonged and pronounced inflammation observed in conditions like sepsis. Yet, very few experimental studies have administered endotoxin using both  strategies, leaving significant gaps in determining the underlying mechanisms responsible for their differing immune responses. We use mathematical modeling to analyze cytokine data from two studies administering a 2 ng/kg dose of endotoxin, one as a bolus and the other as a continuous infusion over four hours. Using our model, we simulate the dynamics of mean and subject-specific cytokine responses as well as the response to long-term endotoxin administration. Cytokine measurements reveal that the bolus injection leads to significantly higher peaks for IL-8, while IL-10 reaches significantly higher peaks during continuous administration. Moreover, the peak timing of all measured cytokines occurs later in the continuous infusion. We identify three model parameters that significantly differ between the two administration methods. Monocyte activation of IL-10 is greater during the continuous infusion, while recovery rates of IL-8 is faster for the bolus injection. This suggests that a continuous infusion elicits a stronger, longer-lasting systemic reaction through increased stimulation of monocyte anti-inflammatory mediator production and decreased recovery of pro-inflammatory catalysts. Furthermore, our continuous infusion model exhibits prolonged inflammation with recurrent peaks resolving within two days during long-term (20-32 hours) endotoxin administration.}\\

\noindent \textbf{{\color{bg}\Large Key Points}}
\begin{itemize}
    \item This study uses mathematical modeling to compare the dynamic response to continuous and bolus endotoxin administration examining the hypothesis that continuous infusion better represents the inflammation seen in clinical scenarios such as sepsis.
    \item Our modeling study augments limited experimental studies by providing a better understanding of the pathways and mechanisms impacting the change in immune responses with long-term endotoxin infusion.
    \item We introduce a  mathematical model incorporating essential cellular and cytokine pathways and calibrate it to mean and subject-specific data from two endotoxin studies.
    \item Statistical analysis of optimized model parameters suggests that the monocyte activation rate for IL-10 is greater for the continuous infusion and degradation rates of TNF-$\alpha$ and IL-8 are greater for the bolus injection.
    \item Continuous infusion model simulations show that prolonged LPS administration (20-32 hours) significantly extends the cytokine response and results in recurrent cytokine spikes. After 32 hours, the stimuli becomes too weak to elicit a system response.
\end{itemize}

\section*{{\color{bg}Introduction}}

Endotoxin (lipopolysaccharide, LPS) derived from gram-negative bacteria's outward membrane \citep{Heine2001} is an immunostimulant administered to healthy subjects as an experimental procedure to study the inflammatory response \citep{Suffredini2014}. This type of experiment, referred to as an endotoxin challenge, has allowed insight into mechanisms and treatments of inflammation events such as rheumatoid arthritis \citep{Miller1979,Merrill2011,Lorenz2013}, systemic lupus erythematosus \citep{Zuckerman1996}, cancer \citep{Easson1998, Ho2013, Yassine2016}, Alzheimer's disease \citep{Sly2001,Cunningham2005,Akimoto2007}, and sepsis \citep{Fitzal2003, Fredriksson2009,Shinozaki2010,Leijte2019}.

In an endotoxin challenge, LPS can be administered as a bolus (instantaneous) injection \citep{Copeland2005,Clodi2008,Janum2016}, a continuous infusion over several hours \citep{Berg2012}, or a combination of the two \citep{Kiers2017} in both humans and animals \citep{Bahador2007}. The response is an increase in pro- (TNF-$\alpha$, IL-1$\beta$, IL-6, IL-8) and anti- (IL-10, IL-1ra) inflammatory cytokines, immune cells, body temperature, heart rate, blood pressure, and hormone levels \citep{Givalois1994,Bahador2007,Clodi2008,Janum2016}. The peak of each measured quantity and the time it takes to reach the peak vary depending on a host of controllable (administration method and total dose administered) and uncontrollable (individual variation due to genetics, sex, and health status) factors. 

\citet{Taudorf2007} performed an endotoxin challenge in healthy men administering 0.3 ng/kg of LPS as a bolus and a continuous infusion over 4 hours. They found that the administration method significantly affects TNF-$\alpha$, IL-6, and neutrophil production rates. These measured quantities peaked earlier and had larger magnitudes during the bolus administration than the continuous infusion. \citet{Kiers2017} compared immune responses to 1 and 2 ng/kg bolus doses of LPS in addition to a 1 ng/kg bolus followed by a 3 ng/kg continuous infusion over 3 hours. This study showed a significant difference in mean cytokine concentrations, flu-like symptoms (headache, nausea, shivering, pain), temperature, and heart rate increases between the bolus-only and the bolus plus continuous infusion dose. Cytokines responses (TNF-$\alpha$, IL-6, IL-8, IL-10) reached significantly higher peak levels, and subjects exhibited prolonged elevated flu-like symptoms during the bolus plus continuous infusion method. These results demonstrate that continuous infusion initiates a more durable and occasionally more pronounced impact on the immune response during the incitement of inflammation. 

Although experimental studies are suitable for investigating the effects of endotoxin administration method, they do not provide insight into why differing dynamics are observed. This is where the power of mathematical modeling of physiological systems can be applied. Simulations with mathematical models can highlight the underlying mechanisms of disease, aid in disease diagnosis, test and validate treatments and predict patient trajectory and mortality. Numerous mathematical models of inflammation have been developed over the last two decades. \citet{Kumar2004,Day2006,Reynolds2006} developed small but novel mathematical models, highlighting their ability to reproduce inflammation scenarios of clinical relevance and potential to predict treatment strategies. Several models built upon this foundation by adding specific immune cells and cytokines activated during the inflammatory response \citep{Chow2005,Roy2007,Foteinou2009,Su2009,Parker2016,Brady2018,Torres2019}. Others created detailed model  incorporating feedback from other physiological entities such as the cardiovascular system, nervous system, the hypothalamic-pituitary-adrenal (HPA) axis, pain perception, and thermal responses \citep{Scheff2010,Foteinou2011,Malek2015,Bangsgaard2017,Dobreva2021, Windoloski2023}. Several models were calibrated to experimental data or validated in specific patients. Some used data from a bolus endotoxin challenge in animals (mice or rats) \citep{Chow2005, Day2006,Roy2007,Parker2016,Torres2019} while others used bolus data from human subjects \citep{Foteinou2009,Scheff2010,Foteinou2011,Malek2015,Bangsgaard2017, Brady2018, Dobreva2021,Windoloski2023}. These studies demonstrate the need for computational inflammation models that (i) utilize experimental data from a continuous infusion of endotoxin and (ii) investigate the mechanisms behind response differences observed during variations in endotoxin administration method. 

Recent experimental studies \citep{Kiers2017, VanLier2019} propose that a continuous endotoxin infusion is more appropriate to study the prolonged system response during systemic inflammation and sepsis. To provide more insight into understanding what immune signaling components are impacted during the switch from a bolus to continuous infusion, we study the inflammatory response to continuous infusion of endotoxin through the lens of a mathematical model. Doing so provides (i) newfound insight into the response differences between a bolus and continuous administration of endotoxin, (ii) a better mathematical representation to study the dynamics of sepsis, and (iii) a better model to investigate treatments of inflammatory conditions since the continuous infusion prolongs the exposure window for treatment testing.

We present a novel inflammatory response mathematical model predicting innate cytokine responses (TNF-$\alpha$, IL-6, IL-8, IL-10) to a 2 ng/kg bolus and continuous infusion over four hours of endotoxin from \citet{Janum2016} and \citet{Berg2012}. The model structure is rigorously explored through sensitivity and identifiability analysis, and parameter estimation calibrates the model to mean and subject-specific cytokine data. We compare each study's cytokine data to characterize larger endotoxin doses than compared in previous literature and develop statistical uncertainty bounds for the optimal mean model. Mechanisms responsible for varying immune dynamics observed in bolus and continuous infusion experimental studies are hypothesized via statistical analysis of optimized model parameters. We propose that the transition from a bolus to continuous infusion impacts physiologically-relevant components related to IL-10 activation by monocytes and TNF-$\alpha$ and IL-8 degradation rates. Moreover, we use our continuous infusion model to investigate the system response to perturbations in infusion duration and total endotoxin dose administered. This illustrates its ability as a clinically-realistic \textit{in silico} model that can simulate prolonged and pronounced responses.

\section*{{\color{bg}Methods}}

\subsection*{{\color{pr}Ethical approval}}
The current work utilized experimental data from two published studies by \citet{Berg2012} and \citet{Janum2016}. The study by \citet{Berg2012} was approved by the Scientific Ethical Committee of Copenhagen and Frederiksberg Municipalities in Denmark. The data from \citet{Berg2012} was made available by Berg (coauthor of this study). The study by \cite{Janum2016} received approval for the experimental protocol by the Regional Committee on Health Research Ethics and the Regional Monitoring Board, and the study followed the protocols listed in the Declaration of Helsinki. Individual data from \citet{Janum2016} was made available by Janum (coauthor of this study) and Mehlsen (coauthor of \citet{Janum2016}). All participants from both studies gave their written and oral consent. 

\subsection*{{\color{pr}Experimental data}}
The experiments by \citet{Berg2012} and \citet{Janum2016} administered the same total dose of endotoxin (2 ng/kg) to healthy study participants, one as a continuous infusion and one as a bolus injection. Mean and subject-specific cytokine data were measured and used to calibrate our mathematical model.

The study from \citet{Berg2012} investigated the effects of an increase in mean arterial pressure on cerebral autoregulation; it included nine healthy male participants aged 21-25. All study participants were subject to physical examination. Data were only included from subjects with normal blood work and cardiovascular markers. Participants did not take medication, had a typical medical history, were non-sedentary, and infection-free at least four weeks before the study. The study by \citet{Janum2016} was designed to investigate the connection between pain and the innate immune system reaction in 20 male athletes aged 18-35. All study participants had a healthy weight, were non-smokers, and had no signs of illness two weeks prior to the study day. Pre-screening activities involved a review of each subject's medical history, a physical examination, and laboratory work. 

In \citet{Berg2012}, participants were subject to a 4-hour continuous infusion of 2 ng/kg (0.5 ng/kg/hr) of endotoxin (Batch G2 B274, US Pharmacopeial Convention, Rockville, MD, USA) administered via an antecubital catheter. In contrast, the study participants in \citet{Janum2016} received a 2 ng/kg bolus endotoxin dose (Lot EC-6, National Institutes of Health, Bethesda, MD, USA) via a peripheral intravenous catheter following two hours of baseline stabilization. In \citet{Berg2012}, blood samples were taken hourly for the first 4 hours following the start of endotoxin administration and 2 hours after completed endotoxin administration. Measurements from \citet{Janum2016} were taken hourly, starting two hours before endotoxin administration and continuing for six hours following administration. To capture peak response, this study analyzed an additional blood sample taken 1.5 hours after LPS administration. \citet{Janum2016} used ELISA (Meso Scale Discovery, Rockville, Maryland, USA) and \citet{Berg2012} used SECTOR Imager 2400 (Meso Scale Diagnostics, Gaithersburg, MD, USA) to determine concentrations of TNF-$\alpha$, IL-6, IL-8, and IL-10.

\begin{figure}[bt]
    \begin{center}
    \includegraphics[width = 6in]{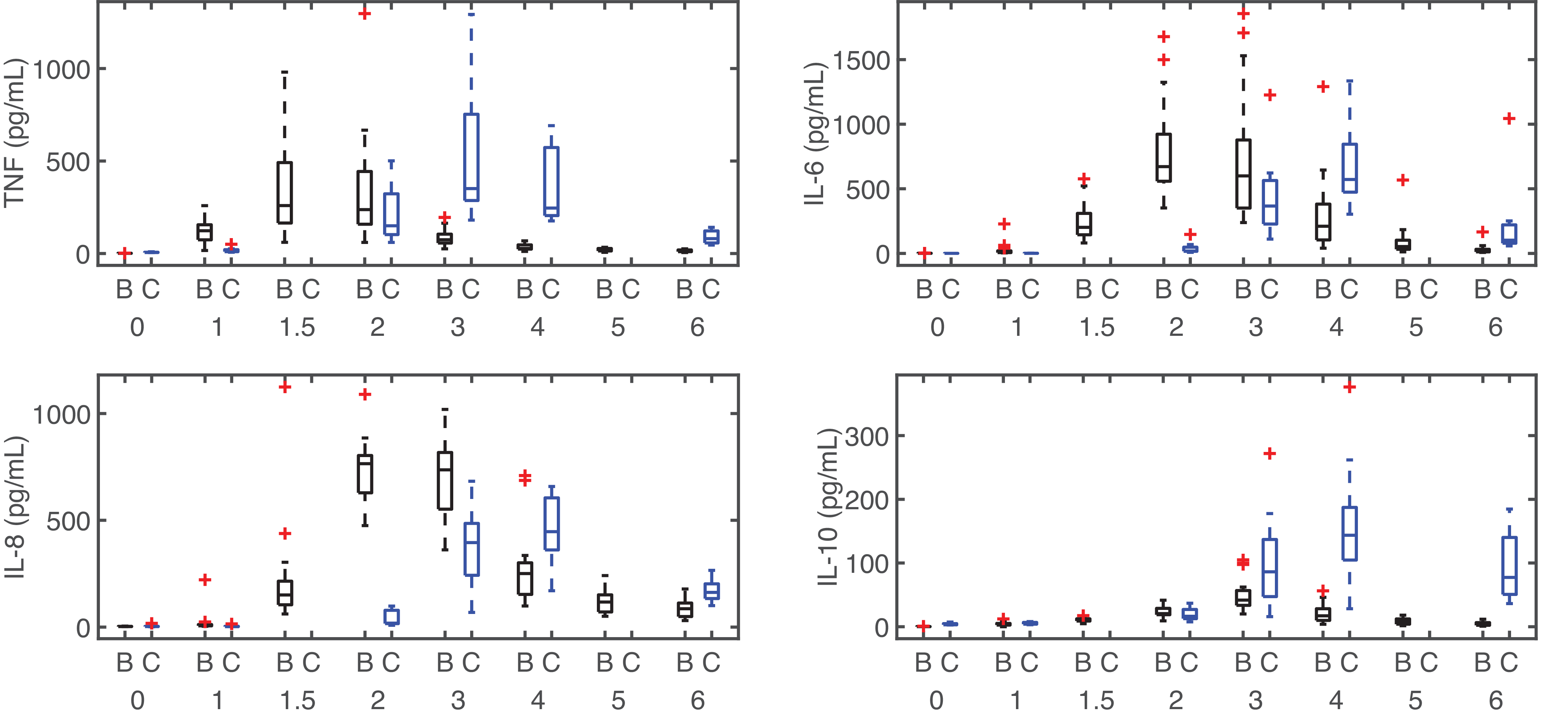}
    \vspace{-0.25cm}
    \end{center}
    \caption{Box and whisker plots of the continuous infusion ($m=9$) \citep{Berg2012} and bolus ($n=20$) \citep{Janum2016} data. The horizontal axis represents the number of hours after the start of endotoxin administration. Black boxplots above the symbol `B' represent bolus data, and blue boxplots above the symbol `C' represent continuous infusion data. 
    The red cross symbol denotes abnormal responses (outliers) from each study. This figure is generated using MATLAB code adapted from \citet{Danz2023}.}
    \label{fig: databoxplots}
\end{figure}

Subjects 1, 2, 6, 8, and 9 from the continuous infusion study were missing one cytokine measurement, and subject 4 was missing four measurements. Of these, subjects 4, 8, and 9 were missing baseline concentration measurements of IL-6. Subject 4 also did not have a baseline concentration of IL-10. Figure \ref{fig: databoxplots} shows data from both studies, identifying outliers from both data sets. Because the immune response exhibits significant variation in individual responses to stimuli, we considered these outlying data points abnormal but not unrealistic.

\begingroup
\renewcommand{\arraystretch}{1.5} 
\begin{table}[bt]
\caption{Experimental data characteristics. Abnormal concentrations (outliers) were not used to compute the mean data. Individual subject concentrations were calculated from a sample size $n=20$ for the bolus data from \cite{Janum2016}, and $m=9$ for the continuous infusion data from \cite{Berg2012}. All concentrations were rounded to the nearest whole number.}
\label{tab:datacharacteristics}
\footnotesize
\centering
\begin{tabular}{cccccc}
\toprule
\textbf{Measurement} & \textbf{Study} & \textbf{TNF-$\alpha$} & \textbf{IL-6} & \textbf{IL-8} & \textbf{IL-10} \\[.15cm] \hline
\multirow{2}{*}{\begin{tabular}[c]{@{}c@{}}
Peak of Mean\\ Concentration (pg/mL)\end{tabular}} & Bolus & 326 & 702 & 714 & 40 \\ 
 & \begin{tabular}[c]{@{}c@{}}  \\ \end{tabular}
Continuous & 532 & 707 & 456 & 136 \\ \hline
\multirow{2}{*}{\begin{tabular}[c]{@{}c@{}}Peak of Subject\\ Concentration (pg/mL)\end{tabular}} & Bolus & 60-1297 & 351-1856 & 522-1124 & 20-105 \\ & 
 Continuous & 212-1293 & 303-1335 & 170-683 & 57-376 \\ \hline
\multirow{2}{*}{\begin{tabular}[c]{@{}c@{}}Peak Timing of Mean\\ Concentration (pg/mL)\end{tabular}} & Bolus & 1.5 & 2 & 2 & 3 \\ & Continuous & 3 & 4 & 4 & 4 \\ \hline
\multirow{2}{*}{\begin{tabular}[c]{@{}c@{}}Peak Timing of Subject\\ Concentration (pg/mL)\end{tabular}} & Bolus & 1.5-2 & 2-3 & 1.5-3 & 2-3 \\ & 
 Continuous & 3-4 & 4 & 3-4 & 3-6 \\ \bottomrule
\end{tabular}
\end{table}
\endgroup

We calculated the mean cytokine response after removing abnormal responses (outlying data points outside the 1.5$\times$IQR range (the 25th and 75th percentile of data) for each endotoxin data set.  Because of the small number of study participants, we only removed abnormal (outlying) measurements instead of that individual's entire cytokine profile. We used the mean of the bolus and continuous infusion data to calibrate our mathematical model. In the remainder of this study, we refer to this as the mean bolus or continuous endotoxin administration. 

Mean and subject-specific cytokine characteristics are reported in Table \ref{tab:datacharacteristics}. Primary pro- and anti-inflammatory cytokines TNF-$\alpha$ and IL-10 had higher mean concentrations during continuous infusion, while secondary cytokines IL-6 did not depend on the administration method, but IL-8 had a higher peak value for bolus injection. With continuous infusion, peak concentrations were later for all measured cytokines. Individual subject concentrations from both studies displayed a considerable variation in cytokine responses; TNF-$\alpha$, IL-6, and IL-8 had a higher variance with bolus injection. 

\subsection*{{\color{pr}Data calibration}}
\begin{figure}[bt!]
    \begin{center}
    \includegraphics[scale=0.25]{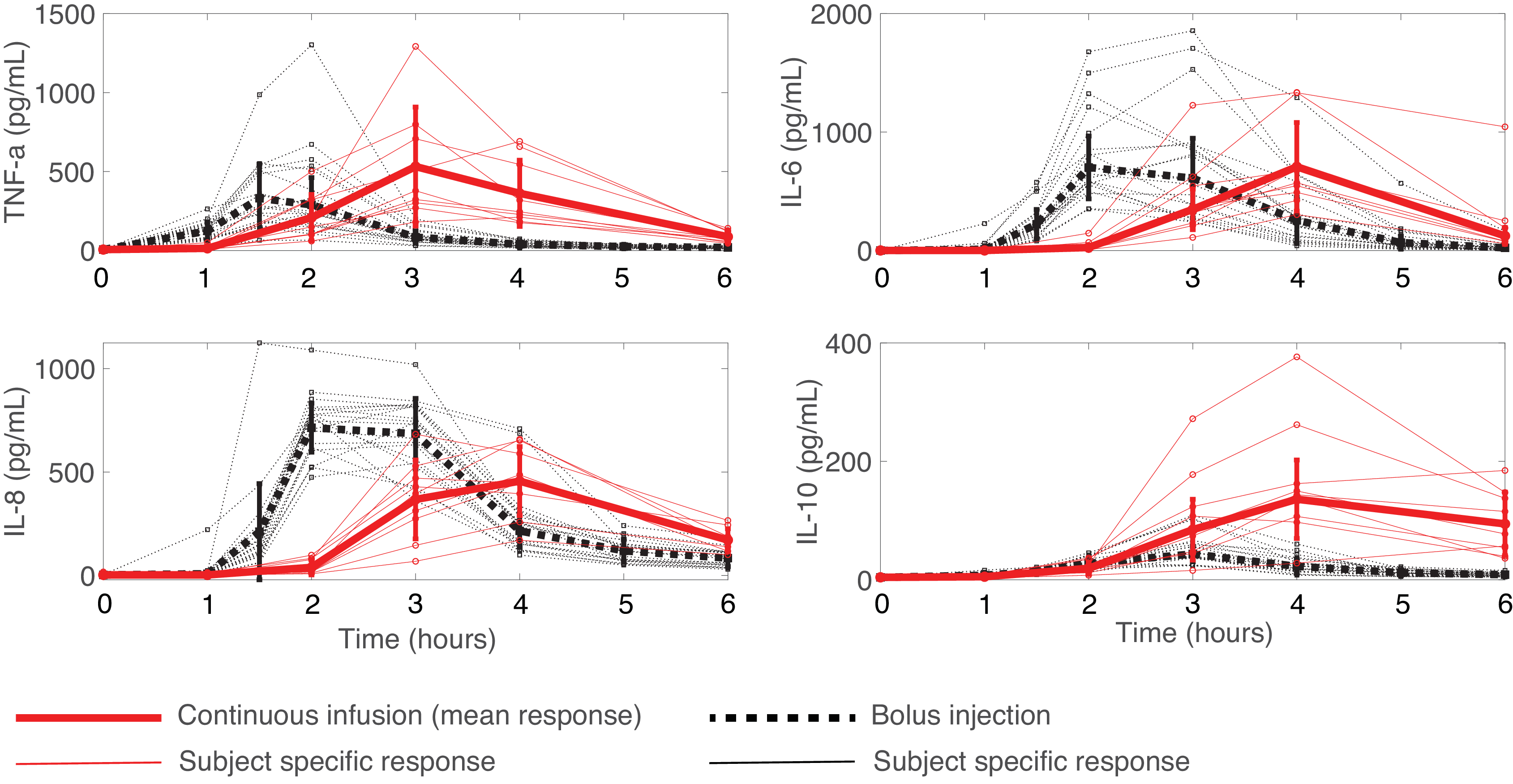}
    \end{center}
    \caption{Mean and subject-specific continuous infusion and bolus data. Red circles denote the continuous infusion data ($m=9$) connected by solid lines, and black squares denote the bolus data ($n=20$) connected by dotted lines. Thin lines represent subject-specific responses and thick vertical lines denote mean (SD).}
    \label{fig:CorrectedData}
\end{figure}

To compare the cytokine responses from the two studies, we adjusted the bolus data so that both studies had the same baseline concentration. This was done by determining the difference ($\bar{d}$) between the mean bolus ($\bar{b}$) and continuous infusion ($\bar{c}$) baseline 
\begin{align*}
    \bar{d}^j_{0} = \bar{b}^j_{0}-\bar{c}^j_{0}
\end{align*}
for cytokine $j=\{TNF,IL6,IL8,IL10\}$ and then shifting the concentrations by 
\begin{align*}
    \hat{b}^j_{i}(k) = b^j_{i}(k)-\bar{d}^j_{0},
\end{align*}
where $b^j_{i}(k)$ is the original cytokine concentration $j$ at time $i$ for the $k$th bolus participant ($1\leq k \leq 20$). The adjusted cytokine concentration is denoted by  $\hat{b}^j_{i}(k)$. Figure \ref{fig:CorrectedData} displays the mean and subject-specific continuous infusion and bolus data.

\subsection*{{\color{pr}Mathematical model}}
\begin{figure}[bt!]
    \begin{center}
    \includegraphics[scale=0.35]{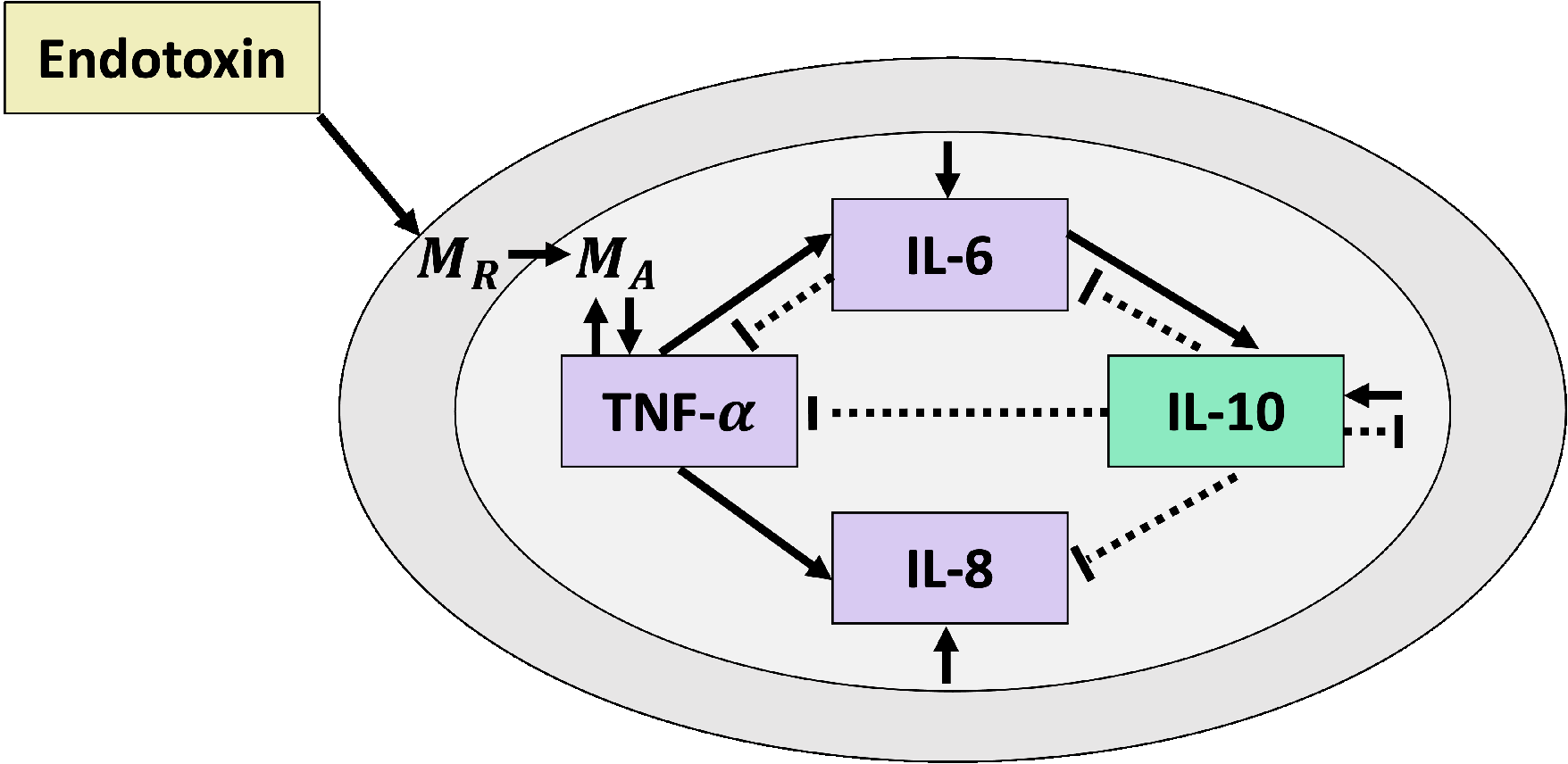}
    \end{center}
    \caption{Cytokine interactions. Endotoxin administration recruits (activates) monocytes from a large pool of resting monocytes ($M_R$). The active monocytes ($M_A$) upregulates production of pro- (TNF-$\alpha$, IL-6, and IL-8) and anti- (IL-10) inflammatory cytokines. TNF-$\alpha$ generates positive feedback on monocyte, IL-6, and IL-8 production. IL-6 exhibits anti-inflammatory properties through self-regulation, downregulation of TNF-$\alpha$, and upregulation of IL-10. IL-10 downregulates all cytokine production and the activation of monocytes. Solid black lines represent stimulation, and dotted black lines represent inhibition.}
    \label{fig:modeldiagram}
\end{figure}

Our mathematical model (Figure \ref{fig:modeldiagram}) adapted from our previous studies \citep{Brady2018,Dobreva2021,Windoloski2023} predicting dynamics of the innate immune response to endotoxin included a system of seven ordinary differential equations (ODEs) with 45 parameters. The equations characterized the time-varying concentrations of endotoxin, resting and activated monocytes, and pro- and anti-inflammatory cytokines. Below we briefly describe the model and refer to \citet{Brady2017} for a detailed derivation of the equations. \\

\noindent{\bf Endotoxin:} The equation determining the endotoxin concentration ($E$, ng/kg) was adapted from \citet{Brady2018,Dobreva2021} to account for a continuous infusion. This formulation is similar to that in \citet{Windoloski2023} and motivated by \citet{Day2006}. The endotoxin rate of change was given by 
\begin{equation}
    \frac{dE}{dt}=\begin{cases}
    D_h-k_EE & t\leq D_{ad}\\
    -k_EE & t>D_{ad}
    \end{cases},
    \label{eq:endotoxin}
\end{equation}
where $D_h$ (ng/kg/hr) was the endotoxin dose administered per hour, $D_{ad}$ (hr) the dosing administration duration, and $k_E$ (hr$^{-1}$) the endotoxin decay rate. The 2 ng/kg continuous infusion was administered over 4 hours so $D_h=0.5$, $D_{ad}=4$, and $E(0)=0$, whereas $D_h=0$, $D_{ad}=0$, and $E(0)=2$ for the bolus injection. \\ 

\noindent{\bf Monocytes:} During the endotoxin challenge, resting monocytes circulating in the blood are activated, upregulating cytokine production. This process regulates inflammation via positive and negative feedback \citep{Rossol2011}. The resting ($M_R$) and activated ($M_A$) monocytes (number of cells, noc) were found from 
\begin{equation}
    \frac{dM_R}{dt} = k_{MR} M_R\left(1-\frac{M_R}{M_\infty}\right) - H_M^U(E)\left(k_M + k_{MTNF} H_M^U(TNF)\right) H_M^D(IL10)M_R \\
    \label{eqMR}
\end{equation}
\begin{equation}
    \frac{dM_A}{dt} =  H_M^U(E)\left(k_M + k_{MTNF} H_M^U(TNF)\right) H_M^D(IL10) M_R - k_{MA} M_A,
    \label{eqMA}
\end{equation}
where $k_{MR}$ (hr$^{-1}$) denoted the regeneration rate and $M_\infty$ (noc) the carrying capacity for the resting monocytes. The activated monocytes were upregulated by endotoxin \citep{Rossol2011} at rate $k_M$ (hr$^{-1}$) and inflammatory cytokine TNF-$\alpha$ at rate $k_{MTNF}$ (hr$^{-1}$). They were also downregulated by anti-inflammatory cytokine IL-10  \citep{Kucharzik1998}. This process was activated relative to the resting monocytes. The increase in activated monocytes caused an identical decrease in the resting monocytes. Finally, the activated monocytes decayed at rate $k_{MA}$ (hr$^{-1}$). 
 
In equations (\ref{eqMR})-(\ref{eqMA}), the upregulation or downregulation of state $Y$ by state $X$ was as
\begin{equation*}
    H^U_{Y}(X)=\frac{X^h}{\eta_{YX}^h+X^h}, \ \ \ 
    H^D_{Y}(X)=\frac{\eta_{YX}^h}{\eta_{YX}^h+X^h},
\end{equation*}
where $h$ represents the steepness of the curve and $\eta_{YX}$ the half-maximum value.\\

\noindent{\bf Inflammatory mediators:} Activated monocytes upregulate cytokines, signaling proteins that promote or suppress inflammation \citep{Janeway2012}. Cytokines that stimulate inflammation, called pro-inflammatory, include TNF-$\alpha$, IL-6, and IL-8 \citep{Johnston2009}. TNF-$\alpha$ is an early pro-inflammatory mediator responsible for the induction of fever \citep{Janeway2012} and recruitment of other pro-inflammatory cytokines \citep{Johnston2009}. IL-6 is a secondary pro-inflammatory mediator primarily involved in the induction of the liver acute phase response \citep{Johnston2009}, although it can exhibit anti-inflammatory properties as well \citep{Tilg1994}. IL-8 is a late pro-inflammatory mediator mainly responsible for recruiting neutrophils to the target site \citep{Bickel1993}. Monocytes also release anti-inflammatory cytokines, particularly IL-10, to counteract pro-inflammatory responses and provide a balanced immune response \citep{Johnston2009}. These four cytokines are essential components of the innate immune response. Their interactions can be predicted by
\begin{equation}
    \frac{dTNF}{dt}=k_{TNFM} H_{TNF}^D(IL6) H_{TNF}^D(IL10) M_A - k_{TNF}(TNF-w_{TNF})\
    \label{eq:TNF}
\end{equation}
\begin{equation}
    \frac{dIL6}{dt} = \left(k_{6M} + k_{6TNF}H_{IL6}^U(TNF)\right)
    H_{IL6}^D(IL6) H_{IL6}^D(IL10) M_A - k_6(IL6-w_{6})
    \label{eq:IL6}
\end{equation}
\begin{equation}
    \frac{dIL8}{dt} = \left(k_{8M} + k_{8TNF}H_{IL8}^U(TNF)\right)
    H_{IL8}^D(IL10) M_A - k_8(IL8-w_{8})
    \label{eq:IL8}
\end{equation}
\begin{equation}
    \frac{dIL10}{dt} =\left(k_{10M}+k_{106}H_{IL10}^U(IL6)\right) M_A - k_{10} (IL10-w_{10}).
    \label{eq:IL10}
\end{equation}
In equation (\ref{eq:TNF}), TNF-$\alpha$ was activated by monocytes \citep{Johnston2009} at rate $k_{TNFM}$ (pg (mL hr noc)$^{-1}$) and downregulated by IL-6 and IL-10 \citep{Tilg1994}. In equation (\ref{eq:IL6}), IL-6 was activated by monocytes and TNF-$\alpha$ \citep{Johnston2009} at rates $k_{6M}$ (pg (mL hr noc)$^{-1}$) and $k_{6TNF}$ (pg (mL hr noc)$^{-1}$), and downregulated by itself \citep{Verboogen2019} and IL-10 \citep{Johnston2009}. Similarly in equation (\ref{eq:IL8}), IL-8 was activated by monocytes and TNF-$\alpha$ at rates $k_{8M}$ (pg (mL hr noc)$^{-1}$) and $k_{8TNF}$ (pg (mL hr noc)$^{-1}$), and downregulated by IL-10 \citep{Johnston2009}. In equation (\ref{eq:IL10}), IL-10 was activated by monocytes and IL-6 at rates $k_{10M}$ (pg (mL hr noc)$^{-1}$) and $k_{106}$ (pg (mL hr noc)$^{-1}$), \citep{Janeway2012,Jin2013}. Cytokines decayed to their baseline concentrations $w_i$ (pg (mL)$^{-1}$) at rate $k_i$ (hr$^{-1}$) for $i= \{TNF,6,8,10\}$.  \\

\noindent{\bf Model summary:} The mathematical model described above was an ODE system of the form
\begin{equation}
    \frac{dX}{dt}=f(t,X;\theta),
    \label{eq:modelsummary}
\end{equation}
where $X\in\mathbb{R}^7$ denoted the time-varying states $X=\{E,M_R,M_A,TNF,IL6,IL8,IL10\}$ determining endotoxin ($E$), monocytes (resting $M_R$ and activated $M_A$), TNF-$\alpha$, IL-6, IL-8, and IL-10 concentrations. The model parameters  $\theta\in\mathbb{R}^{45}$ are listed in Table \ref{tab:pardescription} with subsections indicating what state the parameters belong to. 

To fit the mathematical model to the experimental data, we minimized the least squares cost function, $J$, given by
\begin{equation}
   J=r^Tr, \ \ \ r=[r_{TNF} \hspace{5mm} r_{IL6} \hspace{5mm}  r_{IL8} \hspace{5mm}  r_{IL10}],
   \label{eq:cost}
\end{equation}
where $r_k$ is the residual vector for each state $k=\{TNF, IL6, IL8, IL10\}$ and
\begin{equation}
    r_k=\frac{1}{\sqrt{N}}\left(\frac{\left[y^k_1\hdots y^k_N\right]-y^k_{data}}{\max\left(y^k_{data}\right)}\right).
    \label{eq:optres}
\end{equation}
In equation (\ref{eq:optres}), $N$ refers to the number of data points for each state $k$, $y^k_{i}=g(t_i,X_k(t_i);\theta)$ denotes the model output for the state $k$ at time $t_i$, $1\leq i\leq N$, and $y^k_{data}$ is the associated data. The least squares cost $J$ was minimized using the nonlinear optimization solver, \textit{fmincon}, from MATLAB (MathWorks Inc., Natick, MA, USA). Upper and lower parameter bounds were set by multiplying and dividing the parameters' nominal value by a factor of four. 

\begin{table}[]
\footnotesize
\centering
\begin{tabular}{lllcc}
\toprule \textbf{Parameter} & \textbf{Description} & \textbf{Unit} & \multicolumn{1}{c}{\textbf{\begin{tabular}[c]{@{}c@{}}Nominal \\ Value (C)\end{tabular}}} & \textbf{\begin{tabular}[c]{@{}c@{}}Nominal \\ Value (B) \end{tabular}} \\[0.15cm] \hline
\multicolumn{5}{c}{\textbf{Endotoxin ($E$)}} \\ \hline
\multicolumn{1}{l}{$D_h$} & \multicolumn{1}{l}{$E$ administered pr hour} & \multicolumn{1}{l}{ng (kg  hr)$^{-1}$} & \multicolumn{1}{c}{0.5} & 0 \\
\multicolumn{1}{l}{$D_{ad}$} & \multicolumn{1}{l}{$E$ administered duration} & \multicolumn{1}{l}{hr} & \multicolumn{1}{c}{4} & 0 \\
\multicolumn{1}{l}{$k_E$} & \multicolumn{1}{l}{$E$ decay rate} & \multicolumn{1}{l}{hr$^{-1}$} & \multicolumn{1}{c}{1.01} & 1.01 \\ \hline
\multicolumn{5}{c}{\textbf{Monocytes ($M$)}} \\ \hline
\multicolumn{1}{l}{$k_{MR}$} & \multicolumn{1}{l}{Regeneration rate} & \multicolumn{1}{l}{hr$^{-1}$} & \multicolumn{1}{c}{0.006} & 0.006 \\
\multicolumn{1}{l}{$k_{MA}$} & \multicolumn{1}{l}{Activated decay rate} & \multicolumn{1}{l}{hr$^{-1}$} & \multicolumn{1}{c}{2.51} & 2.51 \\
\multicolumn{1}{l}{$k_{MTNF}$} & \multicolumn{1}{l}{Activation rate by $TNF$} & \multicolumn{1}{l}{hr$^{-1}$} & \multicolumn{1}{c}{9.000} & 8.650 \\
\multicolumn{1}{l}{$k_M$} & \multicolumn{1}{l}{Activation rate by $E$} & \multicolumn{1}{l}{hr$^{-1}$} & \multicolumn{1}{c}{0.041} & 0.041 \\
\multicolumn{1}{l}{$\eta_{ME}$} & \multicolumn{1}{l}{Upregulation half max of $E$} & \multicolumn{1}{l}{ng (kg)$^{-1}$} & \multicolumn{1}{c}{2} & 3.3 \\
\multicolumn{1}{l}{$\eta_{M10}$} & \multicolumn{1}{l}{Downregulation half max of $IL10$} & \multicolumn{1}{l}{pg (mL)$^{-1}$} & \multicolumn{1}{c}{4.394} & 3.884 \\
\multicolumn{1}{l}{$\eta_{MTNF}$} & \multicolumn{1}{l}{Upregulation half max of $TNF$} & \multicolumn{1}{l}{pg (mL)$^{-1}$} & \multicolumn{1}{c}{222.222} & 140.845 \\
\multicolumn{1}{l}{$h_{ME}$} & \multicolumn{1}{l}{Upregulation exp of $E$ on $M$} & \multicolumn{1}{l}{non dim} & \multicolumn{1}{c}{1} & 1 \\
\multicolumn{1}{l}{$h_{M10}$} & \multicolumn{1}{l}{Upregulation exp of $IL10$ on $M$} & \multicolumn{1}{l}{non dim} & \multicolumn{1}{c}{0.3} & 0.3 \\
\multicolumn{1}{l}{$h_{MTNF}$} & \multicolumn{1}{l}{Upregulation exp of TNF on $M$} & \multicolumn{1}{l}{non dim} & \multicolumn{1}{c}{3.16} & 3.16 \\
\multicolumn{1}{l}{$M_\infty$} & \multicolumn{1}{l}{Carrying capacity} & \multicolumn{1}{l}{noc} & \multicolumn{1}{c}{30000} & 30000 \\ \hline
\multicolumn{5}{c}{\textbf{TNF-$\alpha$} ($TNF$)} \\ \hline
\multicolumn{1}{l}{$k_{TNF}$} & \multicolumn{1}{l}{Decay rate} & \multicolumn{1}{l}{hr$^{-1}$} & \multicolumn{1}{c}{0.6} & 1 \\
\multicolumn{1}{l}{$k_{TNFM}$} & \multicolumn{1}{l}{Activation rate by $M$} & \multicolumn{1}{l}{pg (mL hr noc)$^{-1}$} & \multicolumn{1}{c}{1.333} & 0.845 \\
\multicolumn{1}{l}{$\eta_{TNF10}$} & \multicolumn{1}{l}{Downregulation half max of $IL10$} & \multicolumn{1}{l}{pg (mL)$^{-1}$} & \multicolumn{1}{c}{17.576} & 15.536 \\
\multicolumn{1}{l}{$\eta_{TNF6}$} & \multicolumn{1}{l}{Downregulation half max of $IL6$} & \multicolumn{1}{l}{pg (mL)$^{-1}$} & \multicolumn{1}{c}{560} & 560 \\
\multicolumn{1}{l}{$h_{TNF10}$} & \multicolumn{1}{l}{Downregulation exp of $IL10$ on $TNF$} & \multicolumn{1}{l}{non dim} & \multicolumn{1}{c}{3} & 3 \\
\multicolumn{1}{l}{$h_{TNF6}$} & \multicolumn{1}{l}{Upregulation exp of $IL6$ on $TNF$} & \multicolumn{1}{l}{non dim} & \multicolumn{1}{c}{2} & 2 \\
\multicolumn{1}{l}{$w_{TNF}$} & \multicolumn{1}{l}{Baseline concentration} & \multicolumn{1}{l}{pg (mL)$^{-1}$} & \multicolumn{1}{c}{13.873} & 8.793 \\ \hline
\multicolumn{5}{c}{\textbf{IL-6} ($IL6$)}\\ \hline
\multicolumn{1}{l}{$k_6$} & \multicolumn{1}{l}{Decay rate} & \multicolumn{1}{l}{hr$^{-1}$} & \multicolumn{1}{c}{0.66} & 0.66 \\
\multicolumn{1}{l}{$k_{6M}$} & \multicolumn{1}{l}{Activation rate by $M$} & \multicolumn{1}{l}{pg (mL hr noc)$^{-1}$} & \multicolumn{1}{c}{0.81} & 0.81 \\
\multicolumn{1}{l}{$k_{6TNF}$} & \multicolumn{1}{l}{Activation rate by $TNF$} & \multicolumn{1}{l}{pg (mL hr noc)$^{-1}$} & \multicolumn{1}{c}{0.81} & 0.81 \\
\multicolumn{1}{l}{$\eta_{610}$} & \multicolumn{1}{l}{Downregulation half max of $IL10$} & \multicolumn{1}{l}{pg (mL)$^{-1}$} & \multicolumn{1}{c}{35.152} & 31.071 \\
\multicolumn{1}{l}{$\eta_{66}$} & \multicolumn{1}{l}{Downregulation half max of $IL6$} & \multicolumn{1}{l}{pg (mL)$^{-1}$} & \multicolumn{1}{c}{560} & 560 \\
\multicolumn{1}{l}{$\eta_{6TNF}$} & \multicolumn{1}{l}{Upregulation half max of $TNF$} & \multicolumn{1}{l}{pg (mL)$^{-1}$} & \multicolumn{1}{c}{411.11} & 260.563 \\
\multicolumn{1}{l}{$h_{610}$} & \multicolumn{1}{l}{Downregulation exp of $IL10$ on $IL6$} & \multicolumn{1}{l}{non dim} & \multicolumn{1}{c}{1} & 4 \\
\multicolumn{1}{l}{$h_{66}$} & \multicolumn{1}{l}{Downregulation exp of $IL6$ on $IL6$} & \multicolumn{1}{l}{non dim} & \multicolumn{1}{c}{1} & 1 \\
\multicolumn{1}{l}{$h_{6TNF}$} & \multicolumn{1}{l}{Upregulation exp of $TNF$ on $IL6$} & \multicolumn{1}{l}{non dim} & \multicolumn{1}{c}{2} & 2 \\
\multicolumn{1}{l}{$w_6$} & \multicolumn{1}{l}{Baseline concentration } & \multicolumn{1}{l}{pg (mL)$^{-1}$} & \multicolumn{1}{c}{0.610} & 0.610 \\ \hline
\multicolumn{5}{c}{\textbf{IL-8 ($IL8$)}} \\ \hline
\multicolumn{1}{l}{$k_8$} & \multicolumn{1}{l}{Decay rate} & \multicolumn{1}{l}{hr$^{-1}$} & \multicolumn{1}{c}{0.66} & 0.66 \\
\multicolumn{1}{l}{$k_{8M}$} & \multicolumn{1}{l}{Activation rate by $M$} & \multicolumn{1}{l}{pg (mL hr noc)$^{-1}$} & \multicolumn{1}{c}{0.509} & 0.789 \\
\multicolumn{1}{l}{$k_{8TNF}$} & \multicolumn{1}{l}{Activation rate by $TNF$} & \multicolumn{1}{l}{pg (mL hr noc)$^{-1}$} & \multicolumn{1}{c}{0.509} & 0.789 \\
\multicolumn{1}{l}{$\eta_{810}$} & \multicolumn{1}{l}{Downregulation half max of $IL10$} & \multicolumn{1}{l}{pg (mL)$^{-1}$} & \multicolumn{1}{c}{17.576} & 15.536 \\
\multicolumn{1}{l}{$\eta_{8TNF}$} & \multicolumn{1}{l}{Upregulation half max of $TNF$} & \multicolumn{1}{l}{pg (mL)$^{-1}$} & \multicolumn{1}{c}{411.11} & 260.563 \\
\multicolumn{1}{l}{$h_{810}$} & \multicolumn{1}{l}{Downregulation exp of $IL10$ on $IL8$} & \multicolumn{1}{l}{non dim} & \multicolumn{1}{c}{1.5} & 1.5 \\
\multicolumn{1}{l}{$h_{8TNF}$} & \multicolumn{1}{l}{Upregulation exp of $TNF$ on $IL8$} & \multicolumn{1}{l}{non dim} & \multicolumn{1}{c}{3} & 3 \\
\multicolumn{1}{l}{$w_8$} & \multicolumn{1}{l}{Baseline concentration} & \multicolumn{1}{l}{pg (mL)$^{-1}$} & \multicolumn{1}{c}{2.695} & 4.175 \\ \hline
\multicolumn{5}{c}{\textbf{IL-10  ($IL10$)}} \\ \hline
\multicolumn{1}{l}{$k_{10}$} & \multicolumn{1}{l}{Decay rate} & \multicolumn{1}{l}{hr$^{-1}$} & \multicolumn{1}{c}{0.4} & 0.8 \\
\multicolumn{1}{l}{$k_{10M}$} & \multicolumn{1}{l}{Activation rate by $M$} & \multicolumn{1}{l}{pg (mL hr noc)$^{-1}$} & \multicolumn{1}{c}{0.019} & 0.017 \\
\multicolumn{1}{l}{$k_{106}$} & \multicolumn{1}{l}{Activation rate by $IL6$} & \multicolumn{1}{l}{pg (mL hr noc)$^{-1}$} & \multicolumn{1}{c}{0.019} & 0.017 \\
\multicolumn{1}{l}{$\eta_{106}$} & \multicolumn{1}{l}{Upregulation half max of $IL6$} & \multicolumn{1}{l}{pg (mL)$^{-1}$} & \multicolumn{1}{c}{560} & 560 \\
\multicolumn{1}{l}{$h_{106}$} & \multicolumn{1}{l}{Upregulation exp of $IL6$ on $IL10$} & \multicolumn{1}{l}{non dim} & \multicolumn{1}{c}{3.68} & 3.68 \\
\multicolumn{1}{l}{$w_{10}$} & \multicolumn{1}{l}{Baseline concentration } & \multicolumn{1}{l}{pg (mL)$^{-1}$} & \multicolumn{1}{c}{4.239} & 3.747 \\
\bottomrule
\end{tabular}
\caption{Parameter descriptions, units, and nominal values for the continuous infusion (C) and bolus injection (B) endotoxin models. }
\label{tab:pardescription}
\end{table}

\subsection*{{\color{pr}Nominal parameters}}

Nominal parameter values were taken from \citet{Brady2017} except for cytokine baseline concentrations ($w_{TNF}, w_{6},w_{8}$ and $w_{10}$), which were set to the mean continuous infusion and bolus data. We manually adjusted nominal parameters affecting peak timing to account for the observation (Figure \ref{fig: databoxplots}) that the timing of cytokine activation depends on the administration method. To improve the nominal model fit to the peak magnitudes of cytokine profiles, the peak concentration of state $i=\{TNF, IL6, IL8, IL10\}$, denoted $X_i$, was scaled using the technique from \citet{Windoloski2023} where 
\begin{equation}
    X_i=\alpha \Tilde{X_i},
    \label{eq:scaling}
\end{equation}
for the scaling factor $\alpha$ and desired peak concentration $\Tilde{X_i}$. We substituted equation (\ref{eq:scaling}) into the ODE for state $X_i$ giving
\begin{equation*}
    \frac{dX_i}{dt}=f\left(t,\theta,X_i,X_j\right)\hspace{5mm} \implies \hspace{5mm}\frac{d(\alpha \Tilde{X_i})}{dt}= f\left(t,\theta,\alpha \Tilde{X_i},X_j\right)
\end{equation*}
for states $j\neq i$. Therefore,
\begin{equation}
    \frac{d\Tilde{X_i}}{dt}= \frac{1}{\alpha}f\left(t,\theta,\alpha \Tilde{X_i},X_j\right).
\end{equation}
The scaling factor $1/\alpha$ was distributed to each term on the right side of the ODE, scaling the associated parameters in each term. State $X_i$ was also scaled when it was upregulating another state variable, $Y$, as
\begin{equation*}
    H^U_{Y}(X_i)=H^U_{Y}(\alpha \Tilde{X_i})=\frac{(\alpha \Tilde{X_i})^h}{\eta_{YX_i}^h+(\alpha \Tilde{X_i})^h} \hspace{5mm} \implies \hspace{5mm} H^U_{Y}(\Tilde{X_i})=\frac{ \Tilde{X_i}^h}{\left(\frac{\eta_{YX_i}}{\alpha}\right)^h+\Tilde{X_i}^h}.
\end{equation*}
Thus, half-saturation values were scaled by $1/\alpha$. A similar approach was applied for downregulation functions. Baseline cytokine parameters $(w_i)$ were also scaled. Table \ref{tab:pardescription} lists the nominal parameters for the continuous infusion and bolus mean model.

Most nominal parameter values used for model calibration to the individual data were set to the mean optimal values except for the initial cytokine concentrations  ($w_{TNF}, w_{6},w_{8}$ and $w_{10}$), which were set to the individual's cytokine value at baseline. For subjects missing measurements at time zero, we scaled their concentration after one hour based on values from subjects where data were available; IL-6 and IL-10 baseline values were set at $50\%$ and $75\%$ of their concentrations at hour one. The scaling analysis was also applied to the individual subjects because of the significant individual variation in cytokine responses between study participants. To reduce the number of scaled parameters in the subject-specific optimizations, we only scaled cytokines with scaling factor $\alpha<0.9$ or $\alpha>1.1$.

\subsection*{{\color{pr}Sensitivity analysis and subset selection}}

The highly nonlinear mathematical model had seven states and 45 parameters. Because of its structure \citep{Brady2017,Brady2018} and the quantity of data, we selected a parameter subset from the rate constants to estimate. We first conducted a local relative sensitivity analysis as described in \citet{Olufsen2013} on the mean continuous infusion response using the residual vector in equation (\ref{eq:cost}). The sensitivity matrix $\chi$ was given by
\begin{equation}
    \chi=\frac{\partial r}{\partial \log(\theta)}=\frac{\partial y}{\partial \theta}\cdot\frac{\theta}{\max(y_{data})},
    \label{eq:sensmat}
\end{equation}
where $y=g(t,X(t),\theta)$ was the model output at time $t$, $\theta$ the nominal parameter set, and $y_{data}$ the mean continuous infusion data. We approximated the $(i,j)$ entry in the submatrix $\chi_k$ using forward differences, where 
\begin{equation}
    \chi = \left[ \begin{array}{c}\chi_{TNF} \hspace{3mm} \chi_{IL6} \hspace{3mm} \chi_{IL8} \hspace{3mm} \chi_{IL10} \end{array}\right]^T.
\label{eq:chismall}
\end{equation}
For submatrix $X_k$, elements $\chi_{ij}$ were given by
\begin{equation}
\chi_{ij}=\frac{g\left(t_i,X_k(t_i),\theta+he_j\right)-g\left(t_i,X_k(t_i),\theta\right)}{h}\frac{\theta}{\max\left(y^k_{data}\right)},
    \label{eq:chielements}
\end{equation}
where $\phi=10^{-8}$ was the solver tolerance, $h=\sqrt{\phi}$ the step size \citep{pope2009estimation}, and $e_j$ the basis vector in the $j$th direction. We ranked relative sensitivities by computing the two-norm of each column of $\chi$, obtaining a single sensitivity per parameter. We repeated the sensitivity analysis by simulating 100 runs sampling parameters from a uniform distribution varying $\pm 30\%$ around the parameter's nominal value to study effects due to perturbations in parameter values.

Sensitive rate constants were used to select an identifiable parameter subset that can be estimated. We utilized two practical identifiability techniques, the structured correlation method (SCM) and the SVD-QR method \citep{Miao2011, Olufsen2013}. The SCM used the Fisher-information matrix $F=\chi^T \chi$. We checked the condition number to ensure that $F$ had an inverse and calculated $G=F^{-1}$. The matrix $G$ was used to determine the pairwise parameter covariance $C_{ij}$ by
\begin{equation}
    C_{ij}=\frac{G_{ij}}{\sqrt{G_{ii}\  G_{jj}}},
\end{equation}
where $(i,j)$ refers to $\theta_i$ and $\theta_j$. Parameter pairs for which $|C_{ij}|> 0.9$ were considered correlated. The parameter set with the largest correlation was selected. The parameter within that set with the smallest relative sensitivity was removed from the parameter set, and the process was repeated until there were no correlated parameters.

The SVD-QR method used singular value decomposition (SVD) to determine identifiable parameters. This method decomposed the sensitivity matrix $\chi=U\Sigma V^T$ where $U$ and $V$ contained the left and right singular vectors of $\chi$, and $\Sigma$ contained the singular values $\sigma$ of $\chi$. The largest $k$ singular values of $\chi$ were determined by $\sigma(k)\geq 10 \sqrt{\phi}$ where $\phi$ was the ODE solver tolerance. The first $k$ columns of the right singular vectors, $V_k$, were extracted from $V$ and used to find a permutation matrix $P$ such that $V_k^TP=QR$,
where $Q$ was an orthogonal matrix and $R$ was an upper triangular matrix. The permutation matrix $P$ was then used to reorder the parameter vector $\theta$ as
\begin{equation}
    \hat{\theta} = P^T\theta.
\end{equation}
The first $k$ parameters of $\hat{\theta}$ were considered identifiable. 

The subset selection methods determine identifiable parameters near the nominal values. To ensure that optimal values were also identifiable, we investigated the parameter convergence for each subset. We conducted 20 optimizations for each subset with nominal parameters drawn from a uniform distribution of $\pm 10\%$ of each estimated parameter's nominal value. All other parameters were fixed. For each of the 20 runs, we calculated the coefficient of variation ($CoV$) for each estimated parameter $\theta_i$ where \begin{equation}
CoV\left(\theta_i\right)=\frac{\bar{\theta}_i}{\sigma_i}.
\end{equation}
We denoted $\bar{\theta}_i$ as the mean and $\sigma_i$ as the standard deviation of the estimated parameter $\theta_i$. Parameters in each subset with $CoV\geq 0.1$ were identified. The least sensitive parameter was removed from the set, and this process was repeated until all estimated parameters in each subset had a $CoV<0.1$. The resulting parameter subsets were considered sensitive and identifiable and were used for parameter estimation.

\subsection*{{\color{pr}Statistical methods}}

For each parameter subset, the goodness of fit was computed using the coefficient of determination (R$^2$) \citep{dodge2008concise}, the corrected Akaike information criterion (AICc) \citep{burnham2002}, and the Bayesian information criterion (BIC) \citep{schwarz1978}. Details of these measurements are provided in Section 2 of the Supporting Information.

Additionally, we constructed parameter and model confidence and prediction intervals using the frequentist approach detailed in \citet{Seber2003nonlinear,Banks2009,smith2013uncertainty}. Parameter confidence intervals for optimized parameter $\Tilde{\theta}_i$ were computed as 
\begin{equation}
    \Tilde{\theta}_i\pm t_{N-q}^{\alpha/2}\sqrt{\Sigma_{ii}},
    \label{eq:parCI}
\end{equation}
where $N$ was the total number of data points, $q$ was the number of parameters that were estimated, $t_{N-q}^{\alpha/2}$ was the \textit{t}-value from the student's t-distribution for confidence level $1-\alpha$ with $N-q$ degrees of freedom, and the variance estimator matrix $\Sigma$ was given by
\begin{equation}
    \Sigma=(\chi^T(\Tilde{\theta}) V^{-1}\chi(\Tilde{\theta}))^{-1}.
\end{equation}
We defined $\chi$ similarly to equations (\ref{eq:chismall}) and (\ref{eq:chielements}) where 
\begin{equation}
    \chi(\Tilde{\theta}) = \left[ \begin{array}{c}\chi_{TNF}(\Tilde{\theta}) \hspace{3mm} \chi_{IL6}(\Tilde{\theta}) \hspace{3mm} \chi_{IL8}(\Tilde{\theta}) \hspace{3mm} \chi_{IL10}(\Tilde{\theta}) \end{array}\right]^T
\label{eq:chismallUQ}
\end{equation}
and the $(i,j)$ element of submatrix $\chi_k(\Tilde{\theta})$ with $k\in\{TNF,IL6,IL8,IL10\}$ was approximated using forward differences given by
\begin{equation}
    \chi_{ij}(\Tilde{\theta})=\frac{g\left(t_i,X_k(t_i),\Tilde{\theta}+he_j\right)-g\left(t_i,X_k(t_i),\Tilde{\theta}\right)}{h}.
    \label{eq:chielementsUQ}
\end{equation}
Here, $\Tilde{y}^k_i=g(t_i,X_k(t_i),\Tilde{\theta})$ was the optimal model output with optimal parameter vector $\Tilde{\theta}$ for cytokine state $k$ at time $t_i$ for $1\leq i\leq N_k$ where $N_k$ was the number data points for cytokine $k$, and $h$ and $e_j$ were defined as in equation (\ref{eq:chielements}). The diagonal variance matrix $V$ was given by 
\begin{equation}
V=\text{diag}\left(\sigma_{TNF},\sigma_{IL6},\sigma_{IL8},\sigma_{IL10}\right),
\end{equation}
where $\sigma_k$ was a diagonal matrix of size $N_k\times N_k$ with entries
\begin{equation}
    \frac{1}{N_k-q}\left(r_k^Tr_k\right), \ \ \ 
    r_k=\left[\Tilde{y}^k_1\hdots \Tilde{y}^k_{N_k}\right]-y^k_{data}
    \label{eq:parUQres}
\end{equation}
with $y^k_{data}$ defined in equation (\ref{eq:optres}). The asymptotic prediction interval for cytokine $k$ at time $t_i$ was given by
\begin{equation}
    PI^k_i=\Tilde{y}^k_i\pm t_{N_k-q_k}^{\alpha/2}s_k\sqrt{1+\textbf{G}_{ik}^T\left(\chi_k^T(\Tilde{\theta})\chi_k(\Tilde{\theta})\right)^{-1}\textbf{G}_{ik}}
    \label{eq:PI}
\end{equation}
and the confidence interval by
\begin{equation}
    CI^k_i=\Tilde{y}^k_i\pm t_{N_k-q_k}^{\alpha/2}s_k\sqrt{\textbf{G}_{ik}^T\left(\chi_k^T(\Tilde{\theta})\chi_k(\Tilde{\theta})\right)^{-1}\textbf{G}_{ik}}.
    \label{eq:CI}
\end{equation}
We defined $\Tilde{y}^k_i$ and $N_k$ as in equation (\ref{eq:chielementsUQ}), $q_k$ was the number of estimated parameters that impacted cytokine state $k$, and $t_{N_k-q_k}^{\alpha/2}$ was the \textit{t}-value for confidence level $1-\alpha$ with $N_k-q_k$ degrees of freedom. $\chi_k(\Tilde{\theta})$ was given by (\ref{eq:chismallUQ}) and (\ref{eq:chielementsUQ}), but columns in $\chi_{TNF}(\Tilde{\theta})$, $\chi_{IL6}(\Tilde{\theta})$, and $\chi_{IL10}(\Tilde{\theta})$ corresponding to IL-8 parameters were eliminated since they did not impact those state variables (see Figure \ref{fig:modeldiagram}). Entries of these columns were approximately zero and made $F_k=\chi_k^T(\Tilde{\theta})\chi_k(\Tilde{\theta})$ singular unless removed. The matrix $\textbf{G}_{ik}^T$ was defined as
\begin{equation}
    \textbf{G}_{ik}^T=\left( \frac{\partial\Tilde{y}^k_i}{\partial \theta_1} \hdots \frac{\partial\Tilde{y}^k_i}{\partial \theta_{q_k}} \right),
\end{equation}
which was $i$th row of the submatrix $\chi_k(\Tilde{\theta})$, and the variance estimator $s^2_k$ was given by 
\begin{equation}
    s^2_k=\frac{1}{N_k-q_k}r_k^Tr_k,
    \label{eq:varianceestimatorUQ}
\end{equation}
with $r_k$ in equation (\ref{eq:parUQres}). Due to the small number of data points per cytokine, we generated pseudodata to compute uncertainty intervals. Data points at eight and twelve hours were set by quartering the cytokine concentration at six hours and returning the cytokine to the baseline value. Then, a piecewise cubic spline interpolation was performed from $t=0-12$ hours. 

Statistical data analysis included a two-sample unequal variances t-test ($\alpha=0.05$) on the continuous infusion and bolus data before data calibration to compare their maximal concentrations and peak timing statistically. Abnormal cytokine responses (outliers in Figure \ref{fig: databoxplots}) were omitted from the data sampled to conduct the hypothesis test. A two-sample unequal variances t-test ($\alpha=0.05$) was also performed on the set of optimized parameters from subject-specific optimizations to determine statistically significant differences in parameter values between the two administration methods. Parameter values that were outliers within their data set were not included in the data sampled to conduct the hypothesis test.

\section*{{\color{bg}Results}}

\subsection*{{\color{pr} Data}}
Statistical comparison (Table \ref{tab:datapvalues}) of the continuous infusion and bolus injection data show a significantly smaller concentration of IL-8 ($p=0.00147$) and larger concentration of IL-10 ($p=0.00200$) with continuous infusion. The peak concentration for TNF-$\alpha$ ($p=0.0809$) and IL-6 ($p=0.702$) did not statistically differ significantly between the two studies, but the time to peak cytokine concentration was significantly longer for all cytokines during the continuous infusion study: TNF- $\alpha$, IL-6, and IL-8 ($p<0.0001$) and IL-10 ($p=0.00695$). 

\begingroup
\renewcommand{\arraystretch}{1.3} 
\begin{table}[tb!]
\footnotesize
\centering
\begin{tabular}{ccc}
\toprule
\textbf{Cytokine} & \textbf{\begin{tabular}[c]{@{}c@{}}Maximal\\ Concentration\end{tabular}} & \textbf{\begin{tabular}[c]{@{}c@{}}Time to Maximal\\ Concentration\end{tabular}} \\ \hline
TNF-$\alpha$ & $p=0.0809$ ($m=9, n=19$) & $p<0.0001$ ($m=9, n=19$) \\ 
IL-6 & $p=0.702$ ($m=9, n=18$) & $p<0.0001$ ($m=9, n=18$) \\ 
IL-8 & $p=0.00147$ ($m=9, n=19$) & $p<0.0001$ ($m=9, n=19$) \\ 
IL-10 & $p=0.00200$ ($m=8, n=17$) & $p=0.00695$ ($m=8, n=18$)\\ \bottomrule
\end{tabular}
\caption{Statistical significance $(\alpha=0.05$) of data attributes between the continuous infusion ($m$ subjects) and bolus ($n$ subjects) studies in \citet{Berg2012} and \citet{Janum2016}. Subjects with abnormal responses (outliers) for either quantity were omitted from the sample.}
\label{tab:datapvalues}
\end{table}
\endgroup

\subsection*{{\color{pr}Sensitivity analysis and subset selection}}
\begin{figure}[bt!]
    \begin{center}
    \includegraphics[scale=0.25]{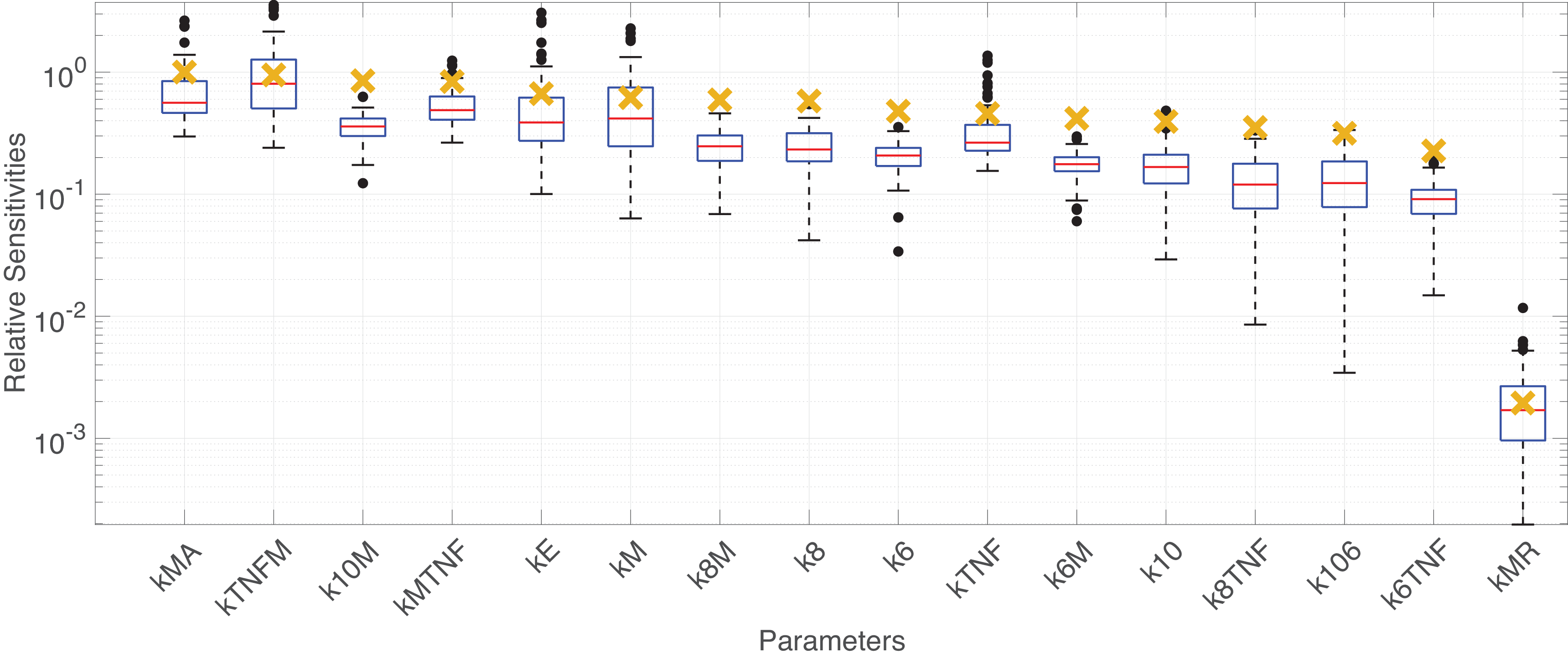}
    \end{center}
    \caption{Ranked sensitivities. Local sensitivities, scaled by the maximum sensitivity,  are denoted by yellow crosses. Boxplots of scaled relative sensitivities are generated from $n=100$ local sensitivity analysis simulations. Values are scaled by the maximum average sensitivity. Black dots denote outliers.}
    \label{fig:sensresults}
\end{figure}

Single and repeated sensitivity analysis (Figure \ref{fig:sensresults}) highlights the system's dependence on endotoxin, activated monocytes, TNF-$\alpha$, and IL-10 dynamics. The system's most sensitive parameters are the growth or decay of these states, where $k_{TNFM}$ (growth rate of TNF-$\alpha$ by monocytes) and $k_{MA}$ (activated monocyte decay rate) have the most significant impact on the model. This can be explained by endotoxin and activated monocytes promoting the activation of cytokines, where TNF-$\alpha$ and IL-10 are the main cytokines that upregulate and downregulate other states. The least sensitive rate constant is $k_{MR}$, the regeneration rate for resting monocytes. Given that our study administers a finite dose of endotoxin that does not deplete the resting monocytes before the system can recover, it is reasonable that this parameter has a minute effect on the system. The single and repeated sensitivity analysis results exhibit similar behavior with minor variations in the order of sensitivity. This observation and careful scaling of nominal parameter values provides a good foundation for choosing identifiable subsets among the sensitive parameters.

The parameter $k_{MR}$ is insensitive, removed from the subset, and fixed at its nominal value. Identifiability analysis using the SCM and SVD-QR method is performed on the remaining 15 sensitive rate constants. Because different identifiability analysis methods are not guaranteed to produce the same results \citep{Brady2017}, we generate three parameter subsets, two using only the SCM and one using the SVD-QR followed by the SCM. Results show that all rate constants for monocytes, IL-6, IL-8, and IL-10 cannot be uniquely estimated. Therefore, the sensitive rate constants are split into two subsets, one that includes monocyte-activated growth rates and the other that contains cytokine-activated growth rates. The SCM performed on each of these subsets results in the two subsets
\begin{align*}
    S_1 &= \{k_E,k_{MA},k_{MTNF},k_{TNF},k_{TNFM},k_6,k_{6TNF},k_8,k_{8TNF}\}\\
    S_2 &= \{k_{MA},k_{M},k_{TNF},k_{TNFM},k_6,k_{6M},k_8,k_{10},k_{10M}\}.
\end{align*}
The third subset is found by performing SVD-QR followed by the SCM on all 15 rate constants, which results in the subset
\begin{align*}
    S_3 &= \{k_{MA},k_{TNF},k_{TNFM},k_6,k_{6M},k_8,k_{8M},k_{10M}\}.
\end{align*}
The identifiability and convergence of the above parameter subsets are checked numerically using the coefficient of variation method, enabling us to reduce the subsets further, obtaining three sensitive and identifiable parameter subsets 
\begin{align*}
    s_1 &= \{k_{MA},k_{MTNF},k_{TNF},k_{TNFM},k_6,k_8,k_{8TNF}\}\\
    s_2 &= \{k_{MA},k_{M},k_{TNF},k_{TNFM},k_6,k_{6M},k_8,k_{10},k_{10M}\}\\
    s_3 &= \{k_{MA},k_{TNF},k_{TNFM},k_8,k_{8M},k_{10M}\}.
\end{align*}
Note that $S_2=s_2$, indicating that the subset $S_2$ was identifiable.

\subsection*{{\color{pr}Parameter estimation and uncertainty quantification}}

Model fit for the mean continuous infusion data (R$^2$, AICc, BIC, and least squares cost $J$) for subsets $s_1, s_2,$ and $s_3$ are reported in Table \ref{table:goodnessoffit_meancts}. Subset $s_3$ has the lowest AICc and BIC values, but the R$^2$ value and least squares cost did not differ significantly between the three subsets. Given the significance of the AICc and BIC values, we conduct the remaining simulations using $S_{Final} = s_3$  including
\begin{align*}
    S_{Final} = \{k_{MA},k_{TNF},k_{TNFM},k_8,k_{8M},k_{10M}\}.
\end{align*}

\begingroup
\renewcommand{\arraystretch}{1.3} 
\begin{table}[bt]
\footnotesize
\centering
\begin{tabular}{cccccl}
\toprule
\textbf{\begin{tabular}[c]{@{}c@{}}Subset\\ Estimated\end{tabular}} &
  \multicolumn{1}{l}{\textbf{\begin{tabular}[c]{@{}l@{}}Number of \\ Parameters\end{tabular}}} &
  \textbf{Average R$^2$} &
  \textbf{AICc} &
  \textbf{BIC} &
  \multicolumn{1}{c}{\textbf{$J$}} \\ \hline
$s_1$ & 7 & 0.915 & 15.4 & 16.6 & 0.0602 \\ 
$s_2$ & 9 & 0.923 & 24.7 & 22.5 & 0.0466 \\ 
$s_3$ & 6 & 0.913 & 11.1 & 13.3 & 0.0547 \\ \bottomrule
\end{tabular}
\caption{Goodness of fit measurements for the optimized subsets $s_1, s_2,$ and $s_3$. The coefficient of determination is denoted as $R^2$, AICc represents the corrected Akaike information criterion, BIC the Bayesian information criterion, and $J$ the least squares cost.}
\label{table:goodnessoffit_meancts}
\end{table}
\endgroup
\begin{figure}[bt]
    \begin{center}
    \includegraphics[width=0.9\textwidth]{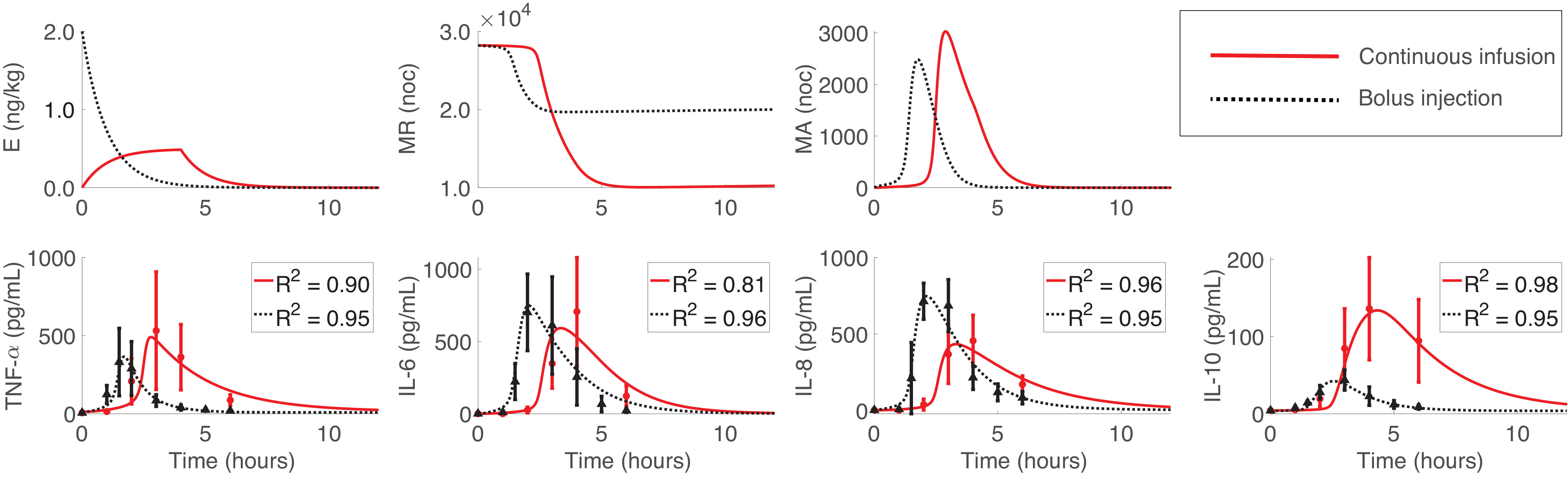}
    \end{center}
    \caption{Optimized model fit to mean continuous infusion and bolus data estimating $S_{Final}$. The mean continuous infusion fit is marked by red solid lines, and the mean (SD) of the data by red circles and error bars. The mean bolus fit has black dotted lines, and black triangles and error bars denote the mean (SD) of the data.}
    \label{fig:optmeanmodel}
\end{figure}

\noindent The mean continuous infusion model exhibits later activation of monocytes and cytokines compared to the bolus injection model (Figure \ref{fig:optmeanmodel}). As a result, the main pro- and anti-inflammatory cytokines TNF-$\alpha$ and IL-10 have larger peak concentrations.  The immune resolution time during the continuous infusion model is approximately ten to twelve hours, whereas the mean bolus model is only six to eight hours. Comparison of model fits by the coefficient of determination (R$^2$) for each cytokine reveal that TNF-$\alpha$ and IL-6 are fitted better by the bolus model, while the continuous infusion model better predicts IL-8 and IL-10. Differences are minor, though, specifically for IL-8 and IL-10.

\begin{figure}[t!]
    \begin{center}
    \includegraphics[width=0.95\textwidth]{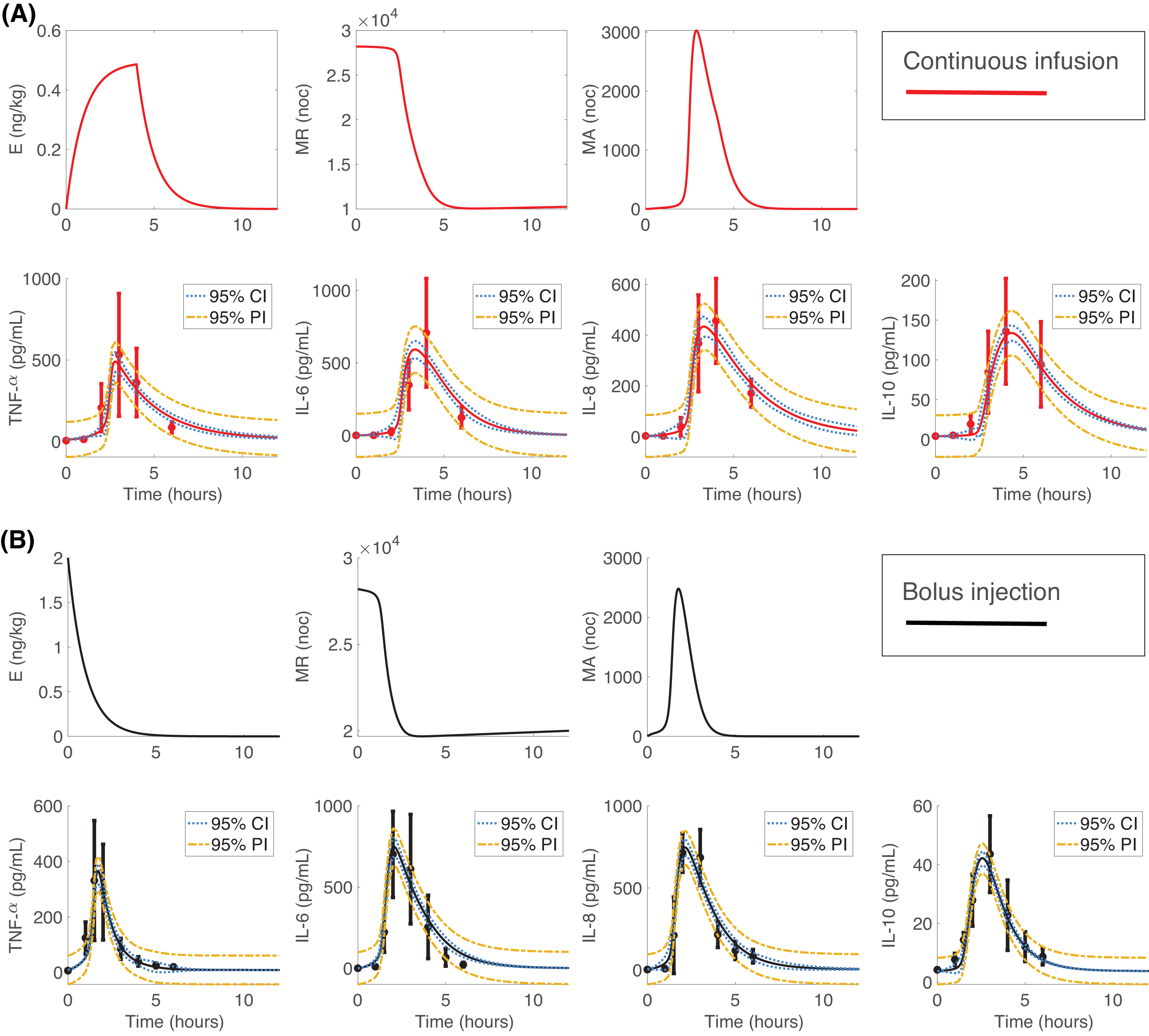}
    \end{center}
    \caption{$95\%$ prediction and confidence intervals for the mean (A) continuous infusion and (B) bolus responses. Mean and SD data points are marked by circles and error bars. Prediction intervals are the yellow dashed-dotted lines, and confidence intervals are the blue dotted lines.}
    \label{fig:UQbounds}
\end{figure}

We generated $N=61$ data points for each cytokine to determine the mean data and model uncertainty using confidence level $(1-\alpha)$ with $\alpha=0.05$. Confidence bounds on the optimal parameters from the continuous infusion and bolus mean model responses are given in Table \ref{tab:UQpars}. The upper and lower bounds remain within the physiological values except for $k_{10M}$, which has a negative lower bound. Prediction and confidence intervals on the optimal mean model are shown in Figure \ref{fig:UQbounds}. Both prediction and confidence intervals for the bolus (Figure \ref{fig:UQbounds}b) are tighter than those for the continuous infusion model (Figure \ref{fig:UQbounds}a), indicating the variability of mean measurements and model output is larger in the continuous infusion data. This is plausible, given the sample sizes of the two studies. The lower bound for the prediction intervals of both dose types extends into negative cytokine values, which is mathematically but not physiologically appropriate.

\begingroup
\renewcommand{\arraystretch}{1.3} 
\begin{table}[t!]
\footnotesize
\centering
\begin{tabular}{ccc}
\toprule
\textbf{Parameter} &
  \textbf{\begin{tabular}[c]{@{}c@{}}Continuous Infusion\\ (Optimal Value $\pm$ Bound)\end{tabular}} &
  \textbf{\begin{tabular}[c]{@{}c@{}}Bolus\\ (Optimal Value $\pm$ Bound)\end{tabular}} \\ \hline
$k_{MA}$   & $3.49\pm 0.0994$  & $2.67\pm 0.0787$ \\ 
$k_{TNF}$  & $0.423\pm 0.132$  & $1.40\pm 0.118$ \\ 
$k_{TNFM}$ & $1.39\pm 0.0696$  & $0.998\pm 0.0398$ \\ 
$k_8$      & $0.386\pm 0.119$  & $0.686\pm 0.100$ \\ 
$k_{8M}$   & $0.613\pm 0.193$  & $0.746\pm 0.163$ \\ 
$k_{10M}$  & $0.0365\pm 0.127$ & $0.0150\pm 0.124$ \\ \bottomrule
\end{tabular}
\caption{Optimal $95\%$ parameter confidence bounds for the mean continuous infusion and bolus model.}
\label{tab:UQpars}
\end{table}
\endgroup

\begin{figure}[h!]
\begin{center}
\includegraphics[width=0.9\textwidth]{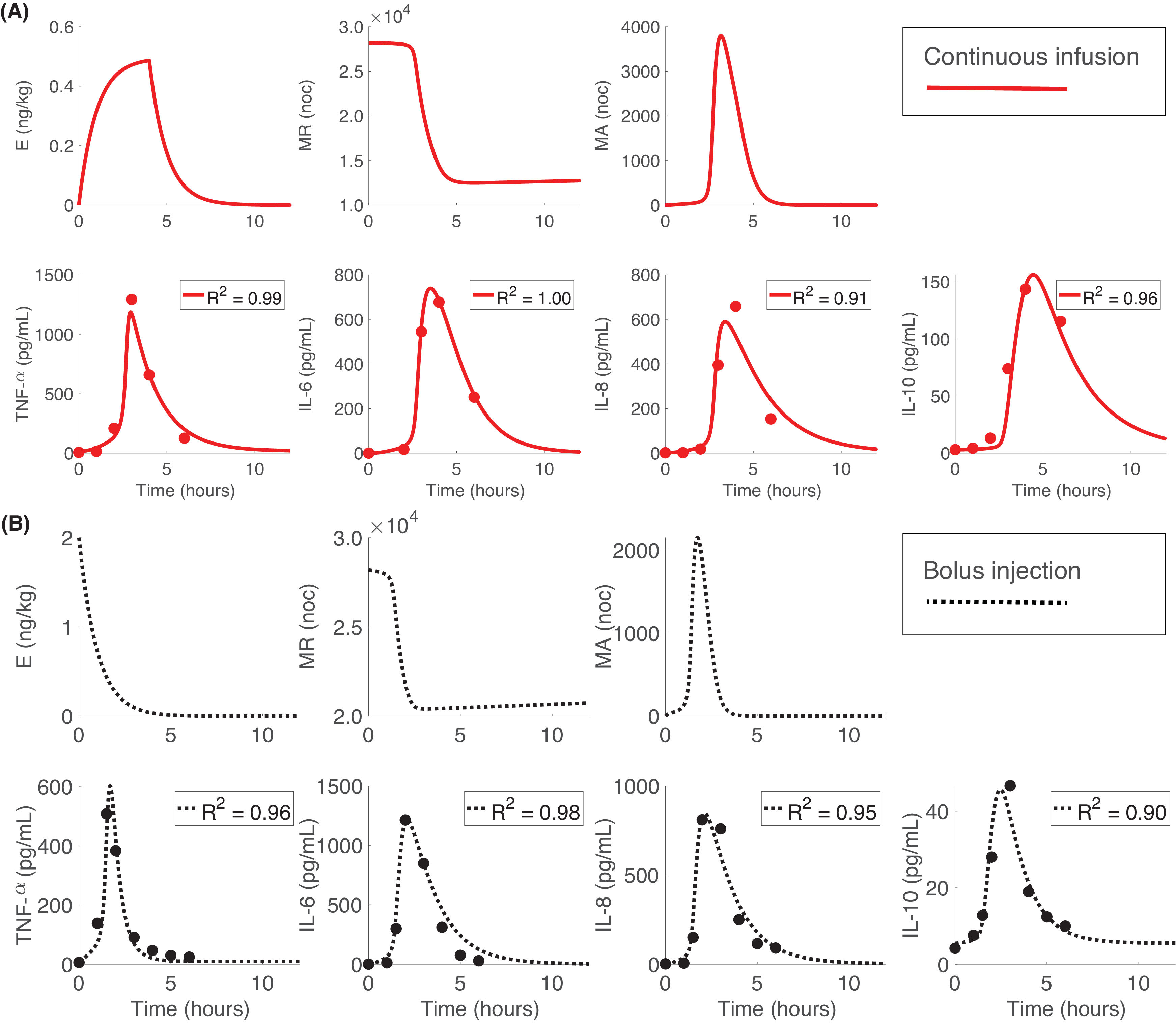}
\caption{Optimal model responses for (A) subject 1 from the continuous infusion study and (B) subject 16 from the bolus study.}
\label{fig:individualmodelfit}
\end{center}
\end{figure}

We fit the model to the subject-specific cytokine profiles from the continuous infusion ($m=9$) and bolus injection ($n=20$) studies by estimating the parameters in $S_{Final}$. Results for continuous infusion subject 1 and bolus injection subject 16  are shown in Figure \ref{fig:individualmodelfit}, and dynamics for the remaining subjects are presented in Figures S1-S29 in the Supporting Information. Results show that our model captures varying cytokine responses to the same total dose of endotoxin for both administration methods. While individual peak cytokine concentrations and peak timing differ from that in the mean response, the model (shown in Figures \ref{fig:individualmodelfit}a and \ref{fig:individualmodelfit}b) is sufficiently robust to capture variation in data. This is evidenced by high R$^2$ values for all subjects.

\begin{figure}[hbt!]
    \begin{center}
    \includegraphics[width=0.9\textwidth]{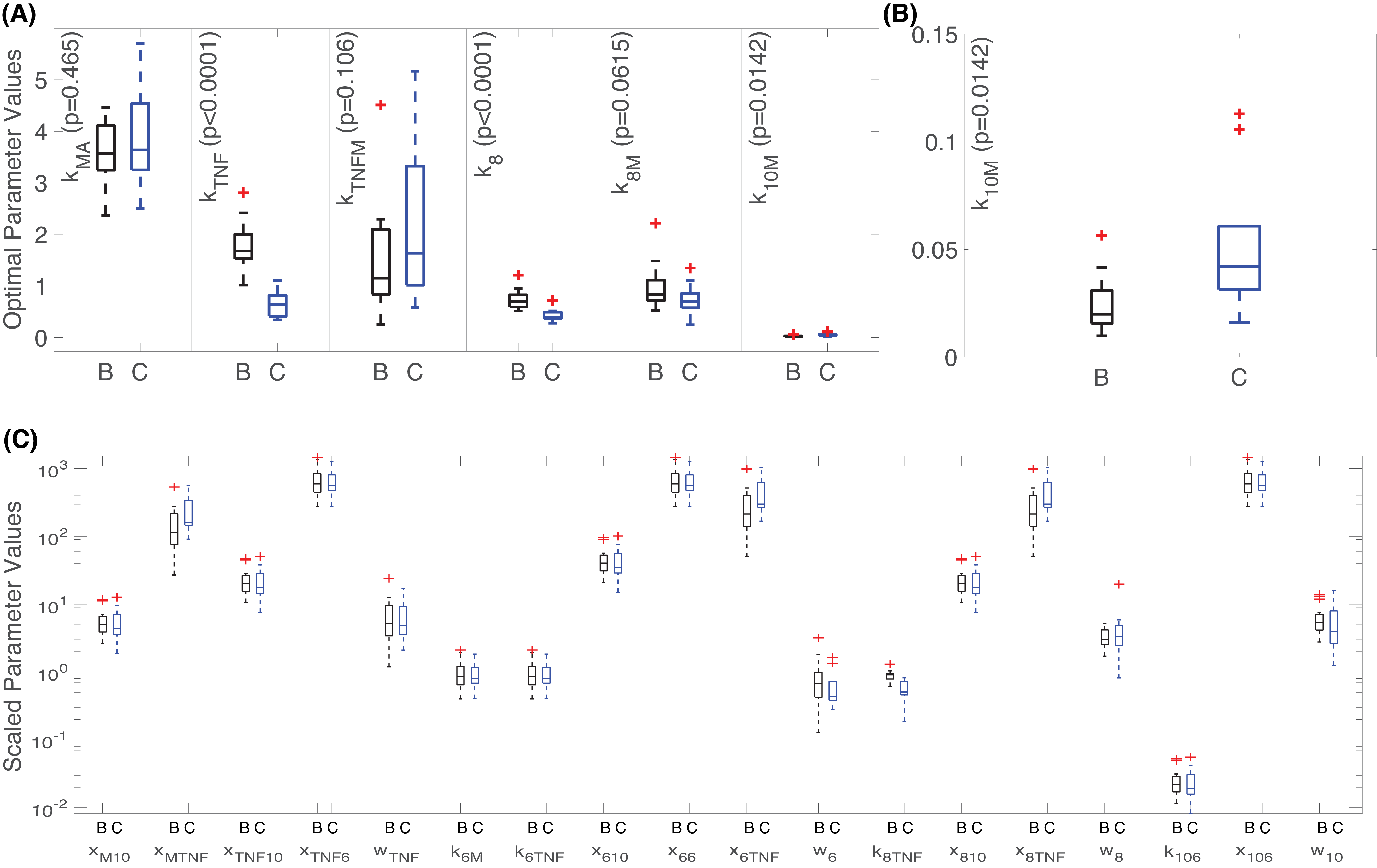}
    \end{center}
    \caption{(A) Boxplots of subject-specific optimized parameters from the bolus `B' (black, $n=20$ subjects) and continuous `C' (blue, $m=9$ subjects) administration models. Associated p-values are listed next to each parameter, (B) zoomed in boxplot of optimized parameter $k_{10M}$ from (A), and (C) boxplots of subject-specific scaled parameters. On all plots outliers are denoted by the red cross. Parameters considered statistically significant ($\alpha=0.05$) include $k_{TNF}$ ($p<0.0001$), $k_{8}$ ($p<0.0001$), and $k_{10M}$ ($p=0.0142$). Parameters not statistically significant include $k_{MA}$ ($p=0.465$), $k_{TNFM}$ ($p=0.106$), and $k_{8M}$ ($p=0.0615$). This figure is generated using MATLAB code adapted from \citet{Danz2023}.}
    \label{fig:parsboxplot}
\end{figure}

The mean and standard deviation for subject-specific optimal and scaled parameters are listed in Table \ref{tab:PSoptpars}, and a boxplot of the optimized subject-specific parameter values are shown in Figures \ref{fig:parsboxplot}A and B. We observe similar median parameter values for the continuous infusion and bolus subject-specific optimizations for parameters $k_{MA}, k_{TNFM},$ and $k_{8M}$. For parameters $k_{MA},k_{TNFM}$, and $k_{10M}$, there is a larger variance in the continuous infusion than the bolus injection. Optimized parameter values denoted as outliers in Figures \ref{fig:parsboxplot}A and B correspond to subjects 3, 5, and 9 from the continuous infusion study and subjects 3, 9, 13, 14, and 20 from the bolus study. These subjects all had abnormal endotoxin responses (at least one outlying data point in Figure \ref{fig: databoxplots}). Boxplots of all scaled subject-specific parameters are shown in Figure \ref{fig:parsboxplot}C and all subject-specific parameter values are listed in Table \ref{tab:PSoptpars}.

Statistical comparison of the continuous infusion and bolus optimized parameters show that $k_{TNF}$ and $k_{8}$ ($p<0.0001$) are significantly larger during the bolus injection, indicating the TNF-$\alpha$ and IL-8 decay faster during the bolus dose. Additionally, $k_{10M}$ ($p=0.0142$) was significantly larger during the continuous infusion, implying that monocyte activation of IL-10 is more pronounced during a continuous infusion of endotoxin. As a result, the continuous infusion had a significantly larger activation response of IL-10 by monocytes and substantially smaller TNF-$\alpha$ and IL-8 degradation rates. Parameters $k_{MA}$ ($p=0.465$), $k_{TNFM}$ ($p=0.106$), and $k_{8M}$ ($p=0.0615$) were not significantly different between the two administration methods, as reported in Table \ref{tab:PSoptpars}. Abnormal responses denoted as outliers in Figure \ref{fig:parsboxplot} were not included in the sample from each study.

\begingroup
\renewcommand{\arraystretch}{1.3} 
\begin{table}[tb!]
\footnotesize
\centering
\begin{tabular}{cccc}
\toprule
\textbf{Parameter} & \textbf{\begin{tabular}[c]{@{}c@{}}Continuous Infusion\\ Mean (SD)\end{tabular}} & \textbf{\begin{tabular}[c]{@{}c@{}}Bolus\\ Mean (SD)\end{tabular}} & \textbf{P-Value} \\ \hline
\boldmath{$k_{MA}$} & \textbf{3.88 (1.06)} & \textbf{3.59 (0.587)} & \boldmath{$p=0.465 (m=9, n=20)$} \\
$\eta_{M10}$ & 5.69 (3.38) & 5.72 (2.79) &  \\ 
$\eta_{MTNF}$ & 249 (152) & 152 (114) &  \\ \hline
\boldmath{$k_{TNF}$} & \textbf{0.642 (0.274)} & \textbf{1.78 (0.406)} & \boldmath{$p<0.0001 (m=9, n=19)$} \\
\boldmath{$k_{TNFM}$} & \textbf{2.25 (1.56)} & \textbf{1.44 (0.948)} & \boldmath{$p=0.106 (m=9, n=19)$} \\
$\eta_{TNF10}$ & 22.8 (13.5) & 22.9 (11.2) &  \\
$\eta_{TNF6}$ & 674 (356) & 686 (346) &  \\ 
$w_{TNF}$ & 7.22 (5.09) & 6.86 (5.22) &  \\ \hline
$k_{6M}$ & 0.975 (0.516) & 0.992 (0.500) &  \\ 
$k_{6TNF}$ & 0.975 (0.516) & 0.992 (0.500) &  \\ 
$\eta_{610}$ & 45.5 (27.1) & 45.8 (22.3) &  \\ 
$\eta_{66}$ & 674 (356) & 686 (346) &  \\ 
$\eta_{6TNF}$ & 461 (280) & 280 (211) &  \\ 
$w_6$ & 0.658 (0.484) & 0.831 (0.690) &  \\ \hline
\boldmath{$k_8$} & \textbf{0.439 (0.127)} & \textbf{0.719 (0.174)} & \boldmath{$p<0.0001 (m=8, n=19)$} \\ 
\boldmath{$k_{8M}$} & \textbf{0.733 (0.331)} & \textbf{0.941 (0.386)} & \boldmath{$p=0.0615 (m=8, n=19)$} \\ 
$k_{8TNF}$ & 0.542 (0.202) & 0.884 (0.154) &  \\ 
$\eta_{810}$ & 22.8 (13.5) & 22.9 (11.2) &  \\ 
$\eta_{8TNF}$ & 461 (280) & 280 (211) &  \\ 
$w_8$ & 5.03 (5.73) & 3.33 (1.07) &  \\ \hline
\boldmath{$k_{10M}$} & \textbf{0.0518 (0.0339)} & \textbf{0.0238 (0.012)} & \boldmath{$p=0.0142 (m=7, n=19)$} \\ 
$k_{106}$ & 0.0250 (0.0148) & 0.0251 (0.012) &  \\ 
$\eta_{106}$ & 674 (356) & 686 (346) &  \\ 
$w_{10}$ & 5.82 (5.04) & 6.280 (3.24) &  \\ \bottomrule
\end{tabular}
\caption{Subject-specific parameter values as mean (SD) and estimated parameter p-values with significance level $\alpha=0.05$. Estimated parameters are marked in bold and remaining parameters were scaled from their nominal values. Mean (SD) values are computed using subject parameter values, including abnormal responses. P-values were calculated by removing the abnormal responses prior to hypothesis testing, with $m$ subjects from the continuous infusion and $n$ subjects from the bolus study being included.}
\label{tab:PSoptpars}
\end{table}
\endgroup

\subsection*{{\color{pr}Infusion perturbations}}
\begin{figure}[tb!]
    \begin{center}
    \includegraphics[width=0.8\textwidth]{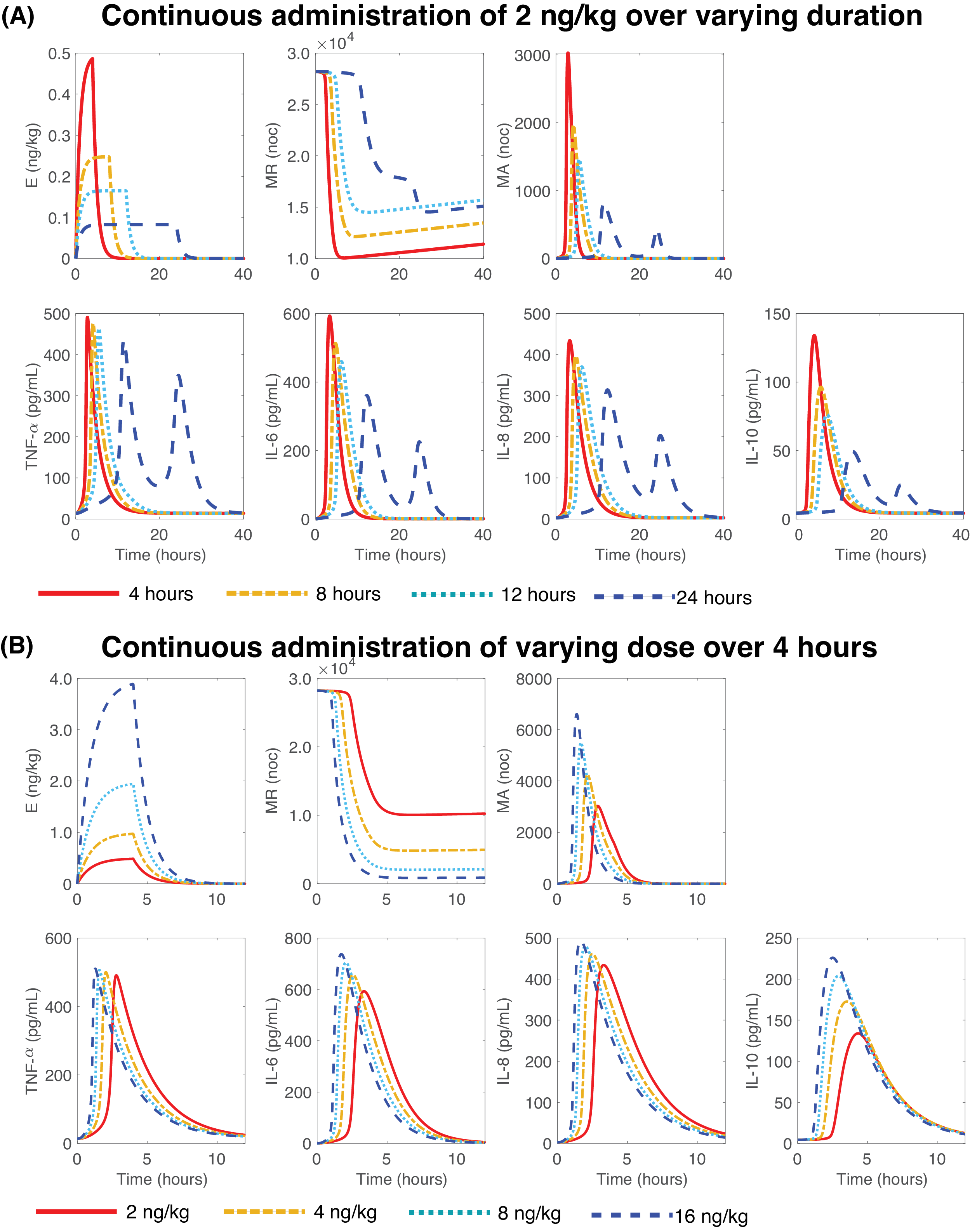}
    \end{center}
    \caption{Continuous infusion mean model response when (A) 2 ng/kg of endotoxin is administered as a 4 (red solid lines), 8 (yellow dashed-dotted lines), 12 (light blue dotted lines), and 24-hour (dark blue dashed lines) continuous infusion, and (B) 2 (red solid lines), 4 (yellow dashed-dotted lines), 8 (light blue dotted lines), and 16 ng/kg (dark blue dashed lines) of endotoxin is administered as a 4-hour continuous infusion.}
    \label{fig:EndotoxinPerturbations}
\end{figure}

We use the optimal mean continuous infusion model to study the response to a longer duration of inflammation and enhanced immune stimulation. Figure \ref{fig:EndotoxinPerturbations}A shows the model response when 2 ng/kg of endotoxin is given continuously over 4, 8, 12, and 24 hours. The infusion duration impacts peak cytokine concentrations and the response's resolution time. Peak concentrations declined and occurred later as the infusion duration increased. Cytokine concentrations returned to baseline approximately 12, 16, 20, and 36 hours following the infusion start for the 4, 8, 12, and 24-hour continuous infusions. The system exhibits oscillatory behavior when the infusion is extended to 24 hours. The increase of anti-inflammatory cytokine IL-10 around 10 hours combats the initial pro-inflammatory response of TNF-$\alpha$ to decline around 12 hours. However, because the endotoxin is still being administered, it rebounds a second time once IL-10 levels begin to decline. Following the termination of endotoxin administration, the monocytes are no longer activated, and as a result, the inflammatory markers return to baseline. This recurrent inflammatory behavior transpires when endotoxin is administered for 20 to 32 hours, after which the stimulation from the endotoxin is not strong enough to induce a pronounced response (Figure S35 in the Supporting Information). This simulation also shows that the system takes approximately 21-23 days to recover (Figure S36 in the Supporting Information) relative to the resting monocyte population returning to the baseline value.

Figure \ref{fig:EndotoxinPerturbations}B displays the model response for a 4-hour continuous endotoxin infusion of 2, 4, 8, and 16 ng/kg. The total endotoxin dose impacts peak cytokine concentrations and the immune resolution time. Larger doses of endotoxin result in earlier, greater peak cytokine concentrations, which occur approximately 1.5-2 hours before peak cytokine concentrations for smaller endotoxin doses. Additional simulations increasing both the duration of the continuous infusion and the total dose of endotoxin are shown in Figures S30-S34 the Supporting Information.

\section*{{\color{bg}Discussion}}

This study uses mathematical modeling to compare bolus and continuous administration of LPS. The model is calibrated to data from \citet{Berg2012} and \citet{Janum2016}. Data analysis reveals that IL-10 has a significantly higher peak for the continuous dose, while the peak IL-8 concentration is higher with the bolus injection.  For the continuous dose, the peaks appear significantly later for all cytokines, a trend that continues when the dose is given over a longer time. Model parameter analysis provide insight into what processes may change with administration methods. Our results suggest that the continuous infusion of endotoxin increases the monocyte production rate of anti-inflammatory cytokine IL-10 and decreases the clearance rates of significant pro-inflammatory markers TNF-$\alpha$ and IL-8. Our continuous infusion model is crucial as it can replicate characteristics of clinical inflammation associated with prolonged elevation of immune markers when endotoxin infusion is extended and pronounced cytokine responses when the endotoxin dosage is increased. Interestingly, administration over 20 and 32 hours produces double cytokine peaks indicating inflammation recurrence.  Finally, we found that it takes over 20 days before the resting monocytes have reached the same level as before the stimulus.

Our findings agree with observations in \citet{Kiers2017} comparing a 2 ng/kg bolus response to a 1 ng/kg bolus followed by a 3 ng/kg continuous infusion. A bolus injection followed by a continuous infusion showed higher IL-10 production and prolonged symptoms due to an extended elevation of cytokines. They also reported a significantly higher production of TNF-$\alpha$, IL-6, and IL-8 with a bolus injection plus continuous infusion. Our study exhibited an increased TNF-$\alpha$ production, though results were not statistically significant, likely due to the small number of subjects and high variance between subjects. IL-8 had significantly lower peaks during the continuous infusion (Figure \ref{fig:optmeanmodel}). Our findings also agree with the partial conclusion of \citet{Taudorf2007}, who reported that (i) the release of cytokines TNF-$\alpha$ and IL-6 occurred significantly later with the continuous infusion compared to a bolus injection and (ii) TNF-$\alpha$ and IL-6 concentrations were significantly larger for the bolus dose. For our study, IL-6 was higher for the bolus dose but again, results were not significant. \citet{Taudorf2007} also reported larger neutrophil concentrations that peaked earlier with  the bolus dose. Our study did not account for neutrophil dynamics, a component that could be added in future studies. Differences in our findings could be a result of low endotoxin dosage in \citet{Taudorf2007} and   unequal total endotoxin dosing in \citet{Kiers2017}. Overall, our findings implicate that continuous stimulation of the system over hours could promote a more significant anti-inflammatory response to counteract prolonged levels of pro-inflammatory cytokines, leading to lower maximal concentrations of secondary cytokines such as IL-8.

A significant contribution of this study is the statistical analysis of optimal parameter distributions between the two administration methods, which has not been examined in earlier works. Results show that the activation rate of IL-10 by monocytes was significantly larger, and the TNF-$\alpha$ and IL-8 decay rates were substantially lower in the continuous infusion versus the bolus injection. These results suggest that continual endotoxin infusion amplifies the monocyte production of IL-10 and dulls the resolution of pro-inflammatory cytokines TNF-$\alpha$ and IL-8. \citet{Kiers2017} indicates that a continuous infusion of endotoxin is a more probable model of prolonged inflammation in conditions like sepsis, where a hyperinflammatory state (often referred to as a cytokine storm) is accompanied by an immunosuppressive phase with elevated anti-inflammation levels \citep{Nedeva2019,Torres2022}. Thus, stimulation by a continuous infusion of endotoxin may exhibit mild but clear signs of a prolonged pro-inflammatory response and a hyperactive anti-inflammatory response, similar to dynamics observed in sepsis and supporting the hypothesis of \citet{Kiers2017}.

Model analysis demonstrated the reliance of dynamics on the endotoxin, monocyte, TNF-$\alpha$, and IL-10 states. These constituents encompass primary elements of the inflammatory response - immune cells that respond to stimuli and the main pro- and anti-inflammatory cytokines that modulate the response strength \citep{Johnston2009}. Thus, it is plausible that these components strongly dictate immune dynamics. It is also reasonable that the least influential component is the monocyte regeneration rate since the challenges analyzed here are from short-lived low-dose endotoxin exposure and the monocyte pool is not depleted prior to endotoxin clearance. However, if simulating a pathogenic insult, we suspect that the influence of this parameter on system dynamics would significantly increase to clear an infection of much greater magnitude than is safely observed in an endotoxin challenge. 

In Figure \ref{fig:parsboxplot}, several parameter values were marked as outliers for continuous infusion and bolus subject-specific model fits. These parameters correspond to subjects from both studies that exhibited abnormal cytokine responses for at least one of the measured cytokines. Given this, we hypothesize that these subjects may experience a more severe response or even enhanced complications to a clinical inflammation event. While this requires further investigation, it could be explored \textit{in silico} by mathematical modeling.

We also explore variations of endotoxin infusion duration and total dose in Figure \ref{fig:EndotoxinPerturbations}. A comparable simulation was conducted in \citet{Windoloski2023} on a bolus endotoxin model where the total dose was increased, representing the administration of more potent immune stimuli that cannot safely be given to humans and the stimuli strength of clinical infection. Both our study and \citet{Windoloski2023} observed enhanced cytokine production. Our simulations extending the continuous infusion duration correspond to the clinical scenario of continual systemic aggravation by inflammatory stimuli. In this case, the model produces up to approximately 36 hours of elevated immune markers depending on the length of the endotoxin infusion. It also displays attributes similar to endotoxin tolerance, a clinical phenomenon related to a reduced response to endotoxin after initial exposure \citep{west2002endotoxin}, through the appearance of multiple decreasing cytokine peaks when continual endotoxin administration is given across 20 to 32 hours. Oscillations in cytokine concentrations also arise for 24 and 36-hour infusions when the total endotoxin dose is increased from 2 ng/kg to 4, 8, and 16 ng/kg (Figures S33-S34 in Supporting Information), showing that the system can produce fluctuating behavior for prolonged periods of inflammation if the stimuli are large enough. These cytokine oscillations that occur could also be clinically-relevant with reference to recurrent infections where the system returns close to baseline before peaking again. In a clinical setting, however, the inflammatory stimulus is a live pathogen whose concentration would also fluctuate, compared to an endotoxin challenge where the administration of endotoxin is constant until cessation of infusion. This model of prolonged inflammation can be used to study inflammatory dynamics over longer periods and test or validate treatments for inflammatory conditions given the longer endotoxin exposure window.

\subsection*{{\color{pr} Limitations}}

A major limitation is that our model is calibrated to two data sets from \citet{Berg2012} and \citet{Janum2016}. Although both administered a total dose of 2 ng/kg of endotoxin and had similar experimental protocols, the endotoxin was sourced from different vendors. There is widespread individual variation in the human immune response,  evidenced by the individual subject data used in this study (Figures \ref{fig: databoxplots} and \ref{fig:CorrectedData}). However, this is not uncommon. It is well-known that immune responses vary between individuals due in part to uncontrollable factors such as genetics, age, sex, seasonal and circadian influences, and environmental effects \citep{Brodin2017}. Therefore, utilizing additional endotoxin challenge data or, more ideally, comparing the immune responses during both endotoxin administration strategies on the same subjects using the same endotoxin batch would yield the best results. Only a few studies administer large endotoxin doses (such as 2 ng/kg as used here) as both a bolus and a continuous infusion. While \citet{Taudorf2007} administers 0.3 ng/kg of endotoxin as a bolus and a continuous infusion, the cytokine concentrations are notably lower than those from a larger dose \citep{Krabbe2001_1,Janum2016} and near or below reported concentrations in septic patients \citep{Casey1993,Wu2009,Berg2012}. \citet{Torres2022} suggests that most patients are likely in the immunosuppressive stage of sepsis upon hospital admittance, so we suspect initial inflammation levels could be higher than reported in sepsis studies. Therefore data from a 2 ng/kg endotoxin challenge likely yield more realistic cytokine concentrations as observed in sepsis and should be used in our study.

Another limitation is that we do not have enough data to validate our endotoxin perturbation results on the continuous infusion model. Experimental data for a continuous infusion of larger endotoxin doses is not seen in literature except in \citet{Kiers2017} (who administers a total of 4 ng/kg of endotoxin) since larger doses of endotoxin are considered unsafe \citep{Bahador2007}. Safety may also play a role in the lack of experimental studies administering endotoxin for an extensive time beyond 4 hours. Another limitation is that, although our mathematical model is highly nonlinear and complex, there are direct elements of the immune response (cells such as macrophages and neutrophils, cytokines such as IL-1$\beta$ and TGF-$\beta$, and signaling pathways such as the NF-$\kappa$B pathway) and other sources of immune regulation (cardiovascular, nerve, hormonal, metabolic) that are not included here. Although \citet{Janum2016} reports that IL-1$\beta$ was measured in the bolus study, its concentrations were not detectable. Our previous work \citep{Dobreva2021} explored interactions of immune, cardiovascular, thermal, and pain responses during a bolus endotoxin challenge, and \citet{Windoloski2023} expanded on that model to include hormonal regulation. Including these additional factors in the model dynamics could provide clearer insight into processes that activate at different speeds or strengths when the endotoxin challenge administration method is varied between a bolus and continuous infusion. A deeper understanding of continuous infusion dynamics could provide a better translational mathematical model of systemic inflammation such as sepsis, encompassing multi-organ dynamics.

\subsection*{{\color{pr}Future work}}
Further investigation of this work involves expanding our study of continuous infusion dynamics to include immune interactions with other systems, such as the cardiovascular and neuroendocrine systems, thermal, pain, and metabolic regulation, building upon the work in \citet{Windoloski2023}. These components are well-known to impact immune response and regulation \citep{Miller2010,Hjemdahl2011, Kenney2014, Janum2016,Varela2018,Dobreva2021}, and studying how a continuous infusion affects these elements can enhance understanding of clinically prolonged inflammation events. Furthermore, while an endotoxin challenge attempts to mimic the dynamics of a clinical-level immune insult, its duration is finite. It cannot simulate the extensive effects of an actual infection or trauma. Therefore, mathematical modeling can extrapolate dynamics from controlled environments to clinical relevance by looking at the impact of age, smoking, diabetes, or cancer on immune responses. Additional future directions of this study focus on transitioning our endotoxin immune response model to a model of sepsis, a life-threatening condition involving hyperactive immune responses and subsequent organ failure that is still not fully understood \citep{Nedeva2019}. Much research has been focused on identifying a universal biomarker and treatment of sepsis, but one has yet to be accepted within the scientific community \citep{cecconi2018sepsis}. However, recent progress has proposed several candidates, including administering vitamin C \citep{Kashiouris2020, Wald2022} to sepsis patients. Adapting our current model to a model of sepsis could help improve our understanding of the mechanisms of sepsis and provide insight into the efficacy of potential sepsis treatments.

\section*{{\color{bg}Conclusion}}

To enhance understanding of potential mechanisms impacting immune responses to endotoxin, we devised a physiologically-based mathematical model simulating mean and subject-specific dynamics in human volunteers exposed to continuous and bolus endotoxin administration. Comparison of subject-specific optimized parameter values revealed significant differences in the monocyte activation rate of IL-10 and recovery rates of pro-inflammatory cytokines TNF-$\alpha$ and IL-8. This suggests that increased IL-10 activation by monocytes and slower recovery rates of pro-inflammatory cytokines could play a role in the more pronounced anti-inflammatory response and smaller secondary cytokine response seen in the continuous infusion data. Additionally, these factors likely influence the system's elongated and more gradual response to the endotoxin, as seen by the statistically significant later peak concentration times of all cytokines during the continuous infusion. Individuals with abnormal cytokine responses also reported statistically outlying optimal parameter values, suggesting their responses to a clinical infection could result in enhanced (outlying) reactions and complications. Simulations of the continuous infusion for a longer duration or increased dose amount display the model's capability to predict immune responses to prolonged inflammation or more potent inflammatory stimuli. Future directions of this work focus on including a whole-body response model to study the immune mechanisms occurring during a continuous infusion and translating this model to study clinically observed inflammation in sepsis patients.

\pagebreak

\bibliographystyle{apalike-comma}
\bibliography{main}

\section*{{\color{bg}Additional Information}}

\subsection*{{\color{pr}Data availability}}
This study utilizes cytokine data from two previously published studies  \citep{Berg2012,Janum2016}. Time series data and mathematical modeling computer simulations for the mean and each subject in both studies are shown in the manuscript, and computer code for this study is available on the GitHub repository: \url{https://github.com/mjcolebank/CDG_NCSU}.

\subsection*{{\color{pr}Competing interests}}
The authors declare that they have no competing interests.

\subsection*{{\color{pr}Author contributions}}
KAW, RMGB, and MSO were responsible for the design of the study and the acquisition, analysis, or interpretation of data. SJ was responsible for the acquisition of data. KAW and MSO were responsible for the mathematical model calibration, simulation, and interpretations. KAW, SJ, RMGB, and MSO were responsible for drafting or critically editing the manuscript. All authors were responsible for approval of the final version of the manuscript. All authors agree to be accountable for all aspects of the work. All persons designated as authors qualify for authorship and all those who qualify for authorship are listed.

\subsection*{{\color{pr}Funding}}
Ronan M.G. Berg and Susanne Janum were supported in part via funding to the Centre for Physical Activity Research (CFAS) receiving suppot from TrygFonden (grants ID 101390 and ID 20045). 

\subsection*{{\color{pr}Acknowledgements}}
We would like to thank Jesper Mehlsen (Rigshospitalet, Denmark) for his contribution to the acquisition of experimental data and Mitchel Colebank (University of California, Irvine) for his guidance on uncertainty quantification methods.

\subsection*{{\color{pr}Keywords}} Inflammation, cytokines, mathematical modeling, data analysis, continuous infusion, endotoxin challenge, administration method

\end{document}